\documentclass[aps,pra,twocolumn,floatfix,superscriptaddress,nobalancelastpage]{revtex4-1}

\usepackage[english]{babel}
\usepackage{amsfonts}
\usepackage{amsmath}
\usepackage{graphicx}
\usepackage{mathtools}
\usepackage[modulo]{lineno}
\usepackage{hyperref}
\usepackage{tcolorbox}
\usepackage{blindtext}
\usepackage[normalem]{ulem}
\usepackage{color}

\newcommand{\ex}[1]{e^{#1}}
\newcommand{\Ket}[1]{\left|#1\right>}
\newcommand{\Bra}[1]{\left<#1\right|}
\newcommand{\BraKet}[2]{\left<#1|#2\right>}
\newcommand{\KetBra}[2]{|#1\rangle\langle#2|}
\newcommand{\trace}[1]{\mathrm{Tr}\left(#1\right)}

\begin{document}


\title{Quantifying the sensitivity to errors in analog quantum simulation}

\author{Pablo M. Poggi}
\email{Corresponding author: ppoggi@unm.edu}
\affiliation{Center for Quantum Information and Control (CQuIC), Department of Physics and Astronomy, University of New Mexico, Albuquerque, New Mexico 87131, USA}

\author{Nathan K. Lysne}
\affiliation{Center for Quantum Information and Control (CQuIC), Wyant College of Optical Sciences, University of Arizona, Tucson, Arizona 85721, USA}

\author{Kevin W. Kuper}
\affiliation{Center for Quantum Information and Control (CQuIC), Wyant College of Optical Sciences, University of Arizona, Tucson, Arizona 85721, USA}

\author{Ivan H. Deutsch}
\affiliation{Center for Quantum Information and Control (CQuIC), Department of Physics and Astronomy, University of New Mexico, Albuquerque, New Mexico 87131, USA}

\author{Poul S. Jessen}
\affiliation{Center for Quantum Information and Control (CQuIC), Wyant College of Optical Sciences, University of Arizona, Tucson, Arizona 85721, USA}

\date{\today}

\begin{abstract}

Quantum simulators are widely seen as one of the most promising near-term applications of quantum technologies. However, it remains unclear to what extent a noisy device can output reliable results in the presence of unavoidable imperfections. Here we propose a framework to characterize the performance of quantum simulators by linking the robustness of measured quantum expectation values to the spectral properties of the output observable, which in turn can be associated with its macroscopic or microscopic character. We show that, under general assumptions and on average over all states, imperfect devices are able to reproduce the dynamics of macroscopic observables accurately, while the relative error in the expectation value of microscopic observables is much larger on average. We experimentally demonstrate the universality of these features in a state-of-the-art quantum simulator and show that the predicted behavior is generic for a highly accurate device, without assuming any detailed knowledge about the nature of the imperfections.

\end{abstract}
\maketitle

\section{Introduction}
In recent years, powerful quantum information processing devices that outperform their classical counterparts have become a real prospect. One of the most recognized potential applications of these technologies, as envisioned by Feynman \cite{feynman1982}, is to efficiently simulate properties of highly correlated quantum systems which are of interest in condensed matter \cite{jaksch1998,hensgens2017}, quantum chemistry \cite{aspuru2005,arguello2019} and high-energy physics \cite{garcia2017,kokail2019}. Important advances in isolating and manipulating quantum systems while maintaining their coherence properties have led to complex quantum devices composed of several tens of qubits \cite{trotzky2012,bernien2017,jurcevic2017,zhang2017_monroe,arute2019}. However, these systems, now routinely referred to as noisy intermediate scale quantum (NISQ) devices \cite{preskill2018}, do  not  meet  the  highly  demanding requirements of fault tolerant, error-corrected  quantum  computers \cite{campbell2017}. 
NISQ processors are intrinsically imperfect analog machines, subject to a continuum of errors in control, background fields and decoherence.
Even as the quality of these devices continues to improve, it is unknown how such imperfections will in general affect the output of these analog quantum simulators (AQSs), and under which circumstances they yield a reliable output~\cite{hauke2012}.\\

In this context, the issue of how imperfections affect the reliability of quantum processors has been studied in many different settings \cite{georgeot2000,hauke2012,sarovar2017}. Particularly, it has recently been observed that extracting information about certain expectation values in noisy devices is a less demanding task than characterizing the full quantum state. This has been studied in the context of dynamical quantum simulators \cite{heyl2019,sieberer2019,lysne2020}, qubit readout and tomography \cite{paini2019,huang2020} and also in terms of algorithm complexity \cite{bravyi2019}.  In this work we establish a general framework to characterize the robustness of AQSs by linking the average sensitivity of expectation values of generic observables to their spectral properties. We do this by showing that these properties characterize the average dependence of expectation values on the quantum state, leading to a quantitative classification of AQS outputs in terms of macroscopic (robust) and microscopic (fragile) observables. We rigorously derive this relation for both static and dynamical models of imperfect quantum simulators.  Crucially, we demonstrate the predictive power of our framework in a real-world quantum simulator based on quantum control of atomic spins \cite{lysne2020}, and show that the imperfections that naturally affect the device lead to errors whose behavior is in excellent agreement with our theoretical findings.

\section{Effect of imperfections in the output of AQS}

Consider a simulator which prepares a quantum system in a state of interest $\Ket{\psi}$ in a $d$-dimensional Hilbert space. We define the output of the device as the expectation value of some observable $\langle A \rangle=\Bra{\psi}A\Ket{\psi}$. Due to imperfect operation of the simulator, however, the system is prepared in a different state $\Ket{\psi_{\rm sim}}$. Our goal is to characterize how the output of the simulator is affected by these imperfections, as a function of the choice of output observable $A$. For this, we define the simulation error,
\begin{equation}
\delta(A)=\Bra{\psi}A\Ket{\psi} - \Bra{\psi_{\rm sim}}A\Ket{\psi_{\rm sim}},
\label{ec:errorA}
\end{equation}

\noindent and consider the perturbed state to be 
\begin{equation}
\Ket{\psi_{\rm sim}}=\mathcal{N}(\gamma)\left(\Ket{\psi} + \gamma \Ket{\psi_\perp}\right),
\end{equation}
\noindent where $\BraKet{\psi}{\psi_\perp}=0$, $\mathcal{N}(\gamma)^2=(1+\gamma^2)^{-1}$ and $\gamma$ quantifies the deviation of the simulated state from the ideal one. In order to assess how the magnitude of the simulation error depends on the  measured observable and not the particular state of the simulator, we consider its average value over all states $\Ket{\psi}$. To perform the average, we consider the Haar measure over random states in Hilbert space \cite{collins2006,bartsch2009}, which we denote as $\overline{(\ldots)}$. Using standard techniques, in Appendix \ref{app:static} we derive the following relation
\begin{equation}
\overline{\delta(A)^2}=\frac{2\gamma^2 \mathcal{N}(\gamma)^2}{d^2-1}\left(\trace{A^2}-\frac{1}{d}\trace{A}^2\right).
\label{ec:errorA_avg}
\end{equation}  
\noindent Similar results can be derived for cases in which the perturbed state is mixed (see Appendix \ref{app:static}). In order to compare the average error for different observables, it is convenient to shift the spectrum of $A$ such that its minimum eigenvalue is zero (excluding the trivial case $A=\mathbb{I}$), which in turn makes $A > 0$. This leaves the error in Eq. (\ref{ec:errorA}) invariant. Furthermore, in order to characterize the magnitude of the error relative to a typical expectation value for different choices of $A$ (similar in spirit to a signal-to-noise measure) we will consider the average \textit{relative} error, defined as
\begin{equation}
\delta_{rel}(A)^2 = \frac{\overline{\delta(A)^2}}{\overline{\langle A \rangle}^2}.
\label{ec:rel_err}
\end{equation}
\noindent Since $\overline{\langle A \rangle}=\frac{1}{d}\trace{A}> 0$, after evaluating Eq. (\ref{ec:rel_err}) we obtain
\begin{equation}
\delta_{rel}(A) = \sqrt{\frac{2d^2}{d^2-1}\left(\frac{\gamma^2}{1+\gamma^2}\right)\left(\trace{\rho_A^2}-\frac{1}{d}\right)},
\label{ec:errorA_avg_rel}
\end{equation}
\noindent where we introduced the operator $\rho_A\equiv A/\trace{A}$ which is a positive, unit trace, Hermitian operator. Eq. (\ref{ec:errorA_avg_rel}) is our first main result. It says that that the degree of robustness of expectation values to imperfections in the quantum state is dictated, \textit{on average}, by  $\eta(A) \equiv \trace{\rho_A^2}$, which we refer to as the \textit{purity} of the observable $A$ in analogy to the usual (state) purity.\\

\begin{figure}[t!]
\centering
\includegraphics[width=0.8\linewidth]{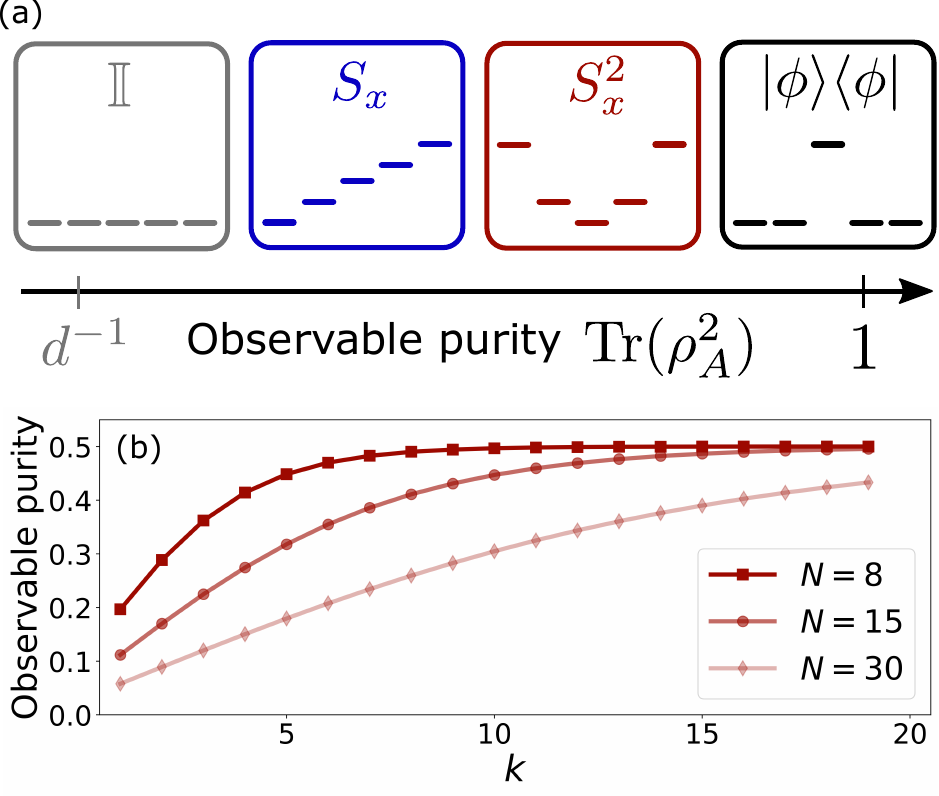}
\caption{{\footnotesize \textbf{(a)} Schematic depiction of the observable purity $\eta(A)=\trace{\rho_A^2}$, where $\rho(A) \equiv A/\mathrm{Tr}(A)$ and its connection to the eigenspectrum of the observable. The spectrum of A is chosen to be nonnegative (see text). Largest purity (equal to $1$) is attained for projectors onto pure states, while lowest purity (equal to $d^{-1}$) corresponds to the identity operator. Intermediate cases correspond to various observables of interest (see main text for details). (b) Purity for even powers of the collective magnetization operator $S_x^{2k}$ with  $S_x=\frac{1}{2}\sum\limits_{i=1}^N \sigma_x^{(i)}$  (direction is chosen arbitrarily).}}
\label{fig:Fig1}
\end{figure}

As depicted in Fig. \ref{fig:Fig1} (a), high purity observables are characterized by having a small, nonextensive set of dominant eigenvalues. As a consequence, their expectation values are greatly affected even by small deviations in the corresponding populations. The extreme case corresponds to projectors onto pure states, i.e., $A=\KetBra{\phi}{\phi}$, which have purity equal to 1, and whose expectation value corresponds to a single state population. On the other hand, small purity observables correspond to high-rank operators, which have an extensive set of eigenvalues close to the mean that contribute to its expectation value, leading naturally to robustness to small deviations in eigenstate populations which tend to average out. The extreme case $\eta(A)=\frac{1}{d}$ is only achieved by $A=\mathbb{I}$, for which the error vanishes trivially. However, many observables of interest show purities which decrease with system size in a similar way. An example of this is given by the collective magnetization in a system of $N$ spin-$\frac{1}{2}$ particles, $S_\alpha=\frac{1}{2}\sum_i \sigma_\alpha^{(i)}$, where $\sigma_\alpha^{(i)}$ denotes Pauli operator acting on the $i$-th particle with $\alpha=x,y,z$. The purity of the collective magnetization evaluates to
\begin{equation}
\trace{\rho_{S_\alpha}^2} = \frac{N+1}{N}2^{-N} \gtrsim 2^{-N} = \trace{\rho_\mathbb{I}^2}.
\end{equation}

\noindent This shows that a physical observable like the magnetization is characterized by an intrinsic robustness to imperfections (on average), which is similar to that of the identity operator for moderately large $N$.\\

The definition of observable purity naturally relates to the notion of typical configurations in statistical mechanics, where observables are taken to be quantities which roughly take the same value over all phase space (compatible with constraints), apart from a small fraction of configurations deemed atypical \cite{review_eth2016}. As a consequence, the value of macroscopic observables is fairly independent of the specific microstate of the system \cite{alvarez2008,bartsch2009,reimann2019}. In our case, Eq. (\ref{ec:errorA_avg_rel}) precisely quantifies robustness of expectation values to deviations in the microstate $\Ket{\psi}$ of the system. This allows us to associate low operator purity with observables that are macroscopic in Hilbert space. Conversely, high purity observables are associated with microscopic quantities that vary sharply, as for example the probability of the system to be in an specific state in an exponentially large state space.\\

Finally, we note that sets of observables with different purities can be constructed in a variety of ways. One example (see Appendix \ref{app:purity} for further analysis) comes from considering powers of a collective spin operator, say $S_x$, whose expectation values can be associated with moments of a probability distribution. In particular, even powers like $S_x^{2k}$ have spectra strongly dominated by degenerate eigenvalues corresponding to the stretched states where all spins are parallel to each other. This is the signature of high purity, as observed in Fig. \ref{fig:Fig1} (b). \\ 

\section{Dynamics of errors in analog quantum simulators}

\subsection{Evolution under weak random perturbations}

\begin{figure*}[!t]
	\centering
	\includegraphics[width=0.95\linewidth]{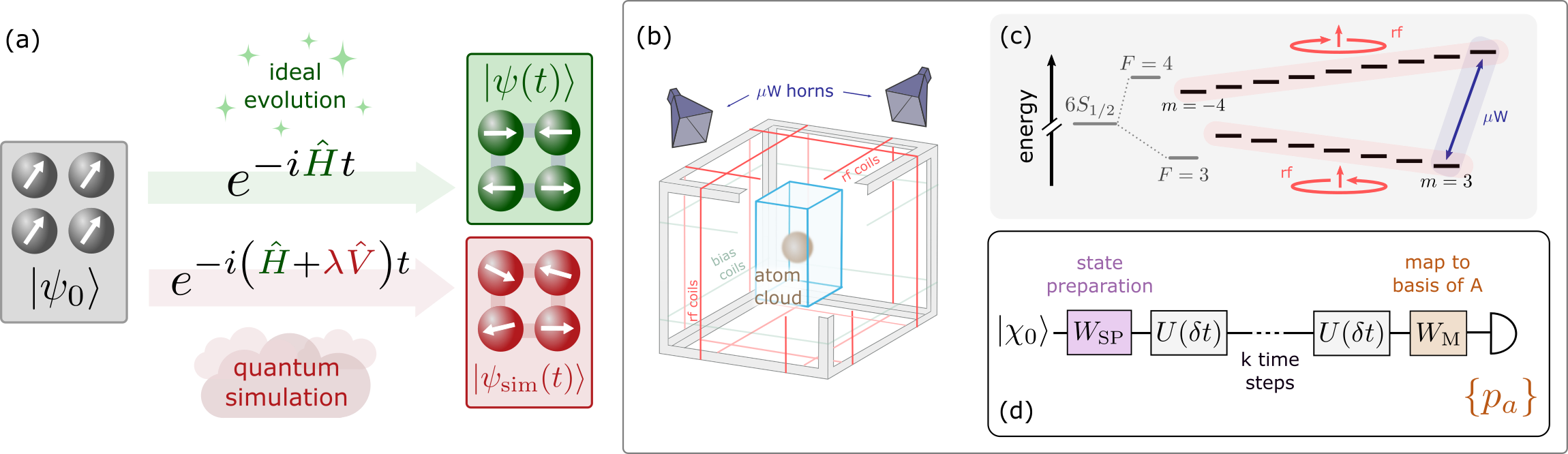}
	\caption{{\footnotesize \textbf{(a)} Schematic model of a dynamical quantum simulator subject to errors. A prepared initial state $\Ket{\psi_0}$ is time-evolved with a Hamiltonian $H$. Ideally, the resulting state is $\Ket{\psi(t)} = \ex{-i H t}\Ket{\psi_0}$. In a real device, different sources of imperfections corrupt the evolution, yielding the perturbed state $\Ket{\psi_{\rm sim}(t)}$. \textbf{(b)} Diagram of the experimental device. \textbf{(c)} Energy level diagram corresponding to the hyperfine ground manifold of cesium, indicating the application of magnetic radiofrequency (rf) and microwave ($\mu$W) control fields. \textbf{(d)} Quantum simulation scheme through discrete time evolution. The unitary maps $W_{\rm SP}$ (state preparation), $U(\delta t)$ (time evolution) and $W_{\rm M}$ (mapping to measurement basis) are generated by the a combination of static and time-dependent control fields, which are programmed using quantum optimal control techniques (see main text for details).}}
	\label{fig:Fig2}
\end{figure*}

A standard protocol for quantum simulation involves engineering a Hamiltonian $H$, under which an initial state $\Ket{\psi_0}$ evolves, leading ideally to $\Ket{\psi(t)}=\ex{-iHt}\Ket{\psi_0}$. Here we analyze this scenario, depicted in Fig. \ref{fig:Fig2} (a) and often referred to as dynamical quantum simulation \cite{zhang2017_monroe,baez2019,trotzky2012}. Assuming that $\Ket{\psi_0}$ can be prepared with high accuracy, then errors will arise in the simulator because of an imperfect implementation of $H$. The nature of such imperfections can be of various kinds, and they depend on the particular physical platform \cite{tacchino2020}.\\

In order to formulate a general model for the impact of errors in AQSs, we will consider that the ideal Hamiltonian dynamics is slightly perturbed in a random way in each run of the simulation, thus leading to an imperfect evolution dictated by a total Hamiltonian $H + \lambda V$, where $\lambda$ is a small dimensionless parameter, and $V$ is a random Hermitian operator characterizing the perturbation. Generalizing Eq. (\ref{ec:errorA}), the error in the output of the dynamical simulator is given by 
\begin{equation}
\delta(A,t)=\Bra{\psi(t)}A\Ket{\psi(t)} - \left[\Bra{\psi_{\rm sim}(t)}A\Ket{\psi_{\rm sim}(t)}\right]_V,
\label{ec:errorAt_def}
\end{equation}
\noindent where $\Ket{\psi_{\rm sim}(t)}=\ex{-i\left(H+\lambda V\right)t}\Ket{\psi_0}$ (here and throughout we set $\hbar=1$) and $\left[\ldots\right]_V$ denotes the average over the random perturbation $V$. Assuming the ideal Hamiltonian $H$ has a nondegenerate spectrum with eigenstates $\left\{\Ket{u_n}\right\}$ and eigenvalues $\left\{E_n\right\}$, and critically, considering $V_{nn}=\Bra{u_n}V\Ket{u_n}$ to be uncorrelated random variables, in Appendix \ref{app:dynamic} we derive an expression for the leading order contribution to the error using standard perturbation theory. The result reads
\begin{equation}
\delta(A,t) = \left(1-f(t)\right)\left[\Bra{\psi(t)}A\Ket{\psi(t)} 
- \trace{\rho_{\psi,D} A}\right]
\label{ec:errorAt}
\end{equation}
\noindent where $\rho_{\psi,D}=\sum_{n} |b_n|^2 \KetBra{u_n}{u_n}$ is the \textit{diagonal ensemble} corresponding to the initial state $\Ket{\psi_0}=\sum_n b_n \Ket{u_n}$ in the eigenbasis of the ideal Hamiltonian $H$ \cite{review_eth2016}. The function $f(t)$ depends on the particular model for the perturbation, and in general obeys $f(0)=1$ and $f(\tau)\rightarrow 0$ for $\tau = \lambda t \gg 1$. The expression obtained in Eq. (\ref{ec:errorAt}) shows that, after a transient time set by the perturbation strength, the simulator error reaches a stationary behavior which depends on the choice of output observable $A$. Results similar to Eq. (\ref{ec:errorAt}) have been obtained even in the nonperturbative regime in the context of thermalization \cite{nation2019,dabelow2020}, giving evidence of the broad validity of the predicted behavior for $\delta(A,t)$.\\

A particular application of Eq. (\ref{ec:errorAt}) is to evaluate it for $A=\rho(t)=\KetBra{\psi(t)}{\psi(t)}$, i.e. the projector onto the unperturbed state of the system. In this case the simulator error in Eq. (\ref{ec:errorAt_def}) equals the infidelity $\mathcal{I}(t)$ or one minus the Loschmidt echo \cite{peres1984,goussev2012}, which measures how well the device simulates the ideal quantum state $\Ket{\psi(t)}$. From Eq. (\ref{ec:errorAt}), we get
\begin{equation}
\delta(\rho(t),t) = \mathcal{I}(t) = (1-f(t))(1-S_0), 
\label{ec:infid}
\end{equation}
\noindent which describes a monotonic increase of the infidelity up to a value $1-S_0$, where $S_0=\trace{\rho_{\psi,D}^2}$ is the inverse participation ratio (IPR) of the state $\Ket{\psi_0}$ in the basis of $H$, as shown originally in \cite{peres1984}.

\subsection{Long-time average error}

The evolution of the fidelity described above is in contrast with the alternative scenario where the output observable is a time-independent operator $A$ (e.g. the magnetization $S_x$). For this case,  the average error given in (\ref{ec:errorAt}) can be thought of as due to deviations from the infinite-time average of $\langle A \rangle $, since
\begin{equation}
\Bra{\psi(t)}A\Ket{\psi(t)} = \sum\limits_{mn} A_{mn} b_m^* b_n\ex{-i(E_n-E_m)t}, 
\end{equation}

\noindent where $b_k = \BraKet{u_k}{\psi}$, and so
\begin{equation}
\lim\limits_{t\rightarrow \infty} \frac{1}{t}\int\limits_{0}^t ds \Bra{\psi(s)}A\Ket{\psi(s)} = \sum\limits_n A_{nn}|b_n|^2 = \trace{\rho_{\psi,D} A}.
\end{equation}

\noindent This implies that for time-independent $A$ we have $\delta(A,t) \rightarrow 0$ on a time average as $t\rightarrow \infty$. {Due to this fact, in the following we focus our discussion on the cumulative error $\mathcal{E}(A,t)$ defined as
\begin{equation}
\mathcal{E}(A,t)^2 = \frac{1}{t}\int_0^t dt' \delta(A,t')^2,
\label{ec:rms_error}
\end{equation}
\noindent which, recalling the definition of $\delta(A,t)$, can be regarded as the root-mean-square error for the expectation value of $A$ over time. 

The asymptotic behavior of $\mathcal{E}(A,t)$ can be easily obtained by inserting 
Eq. (\ref{ec:errorAt}) into the definition of Eq. (\ref{ec:rms_error}), and assuming that the evolution time is large, $\lambda t \gg 1$. The result yields
\begin{equation}
\lim\limits_{t\rightarrow \infty}\mathcal{E}(A,t)^2 = \sum\limits_{n\neq m} |b_n|^2 |b_m|^2 A_{nm} A_{mn}
\label{ec:abserr_inf}
\end{equation}

\noindent where $A_{nm}=\Bra{u_n}A\Ket{u_m}$. As was done in the previous section, in order to extract the dependence of the error on the output observable, we perform the average of Eq. (\ref{ec:rms_error}) over Haar-random initial states and consider the time-dependent average relative error,
\begin{equation}
\mathcal{E}_{rel}(A,t)^2 = \frac{\overline{\mathcal{E}(A,t)^2}}{\overline{\langle A \rangle}^2},
\label{ec:error_rel_At}
\end{equation}
\noindent and denote its asymptotic value for $t\rightarrow \infty$ as $\mathcal{E}^\infty_{rel}(A)$. This quantity generalizes the average relative error $\delta_{rel}(A)$ of Eq. (\ref{ec:errorA_avg_rel}) to the dynamical quantum simulation scheme. Using the techniques discussed in Appendix \ref{app:static} it is straightforward to show that $\overline{|b_n|^2 |b_m|^2}=\frac{1}{d(d+1)}$ for $n\neq m$, and thus combining Eqs. (\ref{ec:abserr_inf}) and (\ref{ec:error_rel_At}) leads to the result

\begin{equation}
\mathcal{E}^\infty_{rel}(A) = \sqrt{\frac{d}{d+1}\left(\trace{\rho_A^2}  - \trace{\rho_{A_D}^2}\right)}.
\label{ec:err_inf}
\end{equation}
\noindent Here $\rho_X = X/\trace{X}$ for both operators $A$ and $A_D=\sum_n A_{nn} \KetBra{u_n}{u_n}$, where $A_{nn}=\Bra{u_n}A\Ket{u_n}$ (recall that $A> 0$ by construction).  Note that the second term inside the square root in Eq. (\ref{ec:err_inf}) depends on the Hamiltonian through its eigenbasis $\{\Ket{u_n}\}$ and can be thought of as Hamiltonian-dependent observable purity
\begin{equation}
\trace{\rho_{A_D}^2} = \frac{\sum A_{nn}^2}{\left(\sum{A_{nn}}\right)^2}
\label{ec:met_ipr_obs}
\end{equation}

\noindent Given the resemblance to the IPR $S_0=\trace{\rho_{\psi_D}^2}$, we can regard Eq. (\ref{ec:met_ipr_obs}) as a measure of the spread of $A$ in the basis of the Hamiltonian. The maximum value of $\trace{\rho_{A_D}^2}$ is $\trace{\rho_A^2}$ , which happens only if $A$ and $H$ commute and are thus diagonal in the same basis. In this regime, the error Eq.  (\ref{ec:err_inf}) vanishes and we need to increase our expansion to the next order in perturbation theory. Of particular interest is when $A$ and $H$ are highly noncommuting, such that $A$ evolves nontrivially under the action of the Hamiltonian. Then, we can then expect $A_{nn}\sim \frac{1}{d}\trace{A}$ and thus $\trace{\rho_{A_D}^2} \sim \frac{1}{d}$. In this generic scenario we then expect 
\begin{equation}
\mathcal{E}^\infty_{rel}(A) \simeq \sqrt{\frac{d}{d+1}\left(\trace{\rho_A^2}  - \frac{1}{d}\right)}
\end{equation}
\noindent which shows the same dependence on the observable $A$ as Eq. (\ref{ec:errorA_avg_rel}). Eq. (\ref{ec:err_inf}) is the main finding of this work, demonstrating that the observable purity $\eta(A)=\trace{\rho_A^2}$ determines the average sensitivity to imperfections in dynamical quantum simulations.

\begin{figure*}[!t]
	\centering
	\includegraphics[width=1.0\linewidth]{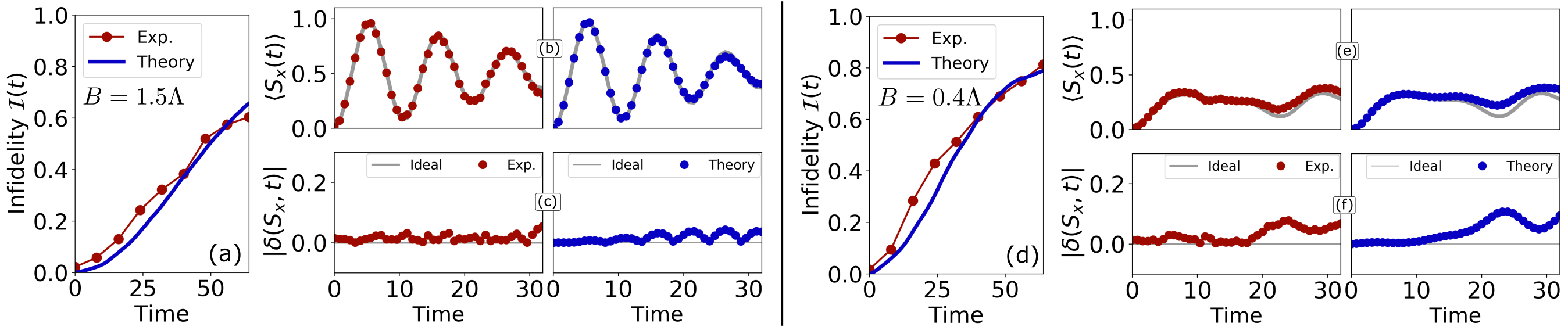}
	\caption{{\footnotesize Comparison between experimental quantum simulation results and the proposed model for dynamics of error generation. Results are shown for the LMG Hamiltonian of Eq. (\ref{ec:lmg_hami}) using \textbf{(a)-(c)} $B=1.5\Lambda$ and \textbf{(d)-(f)} $B=0.4\Lambda$, and starting from the initial state $\Ket{\psi_0}=\Ket{\downarrow_x}^{\otimes N}$. Plots \textbf{(a)} and \textbf{(d)} show state infidelity as a function of simulated time, where red dots correspond to experimental results and continuous blue line to numerical results obtained from the theoretical model, c.f. Eq. (\ref{ec:infid}). The only fitting parameter is the perturbation strength $\lambda$, which is obtained assuming the perturbation $V$ is a Gaussian random matrix and using the procedure outlined in Appendix \ref{app:experimental}. Plots \textbf{(b)} and \textbf{(e)} show dynamics of expectation values for experiment (red) and theory (blue). The theoretical curves are obtained numerically from the model outlined in Sect. III, where the ideal Hamiltonian is perturbed by a random matrix $V$; results shown here are averaged over 50 random instances (see main text for details). Full gray line is the ideal target evolution without errors. \textbf{(c)} and \textbf{(f)} show the time-dependent error in the expectation values $\lvert\delta(A,t)\rvert$ computed for these cases.  Simulated time has units of $1/\Lambda$. Experimental data in panel \textbf{(c)} is also shown in Ref. \cite{lysne2020}.}}
	\label{fig:Fig3}
\end{figure*}

\section{Experimental quantum simulation results}

In the previous section we presented a theoretical framework which relates the magnitude of errors in expectation values with the spectral properties of the corresponding observables. In the following we show that this framework can predict the behavior of such errors in a real-world device, even when no detailed information about the underlying physical imperfections is used in the model. For the work presented here, we use a small, highly accurate quantum (SHAQ) simulator whose state-of-the-art fidelity is critical for the quantitative examination of errors and their impact, but similar studies can in principle be done using a wide range of quantum simulators.


On our experimental platform, previously introduced in \cite{lysne2020}, information is encoded in the hyperfine ground manifold of individual cesium atoms. This space is spanned by the logical basis $\left\{\Ket{F,m}\right\}$ labeled by hyperfine spin quantum numbers $F=3,4$ and $-F\leq m \leq F$, comprising a 16-dimensional Hilbert space. Full unitary controllability over this space is achieved with a combination of a static magnetic bias field along $z$, a pair of phase modulated radio-frequency magnetic fields along $x$ and $y$, and a single phase-modulated microwave magnetic field. As a result, this simulator is universal and can be programmed using quantum optimal control (QOC) to access the dynamics of any Hamiltonian of interest, as in other quantum processors \cite{chow2014,arute2019}. In our simulation scheme, schematically depicted in Fig. \ref{fig:Fig2} (c), the control fields are programmed using QOC techniques to generate the unitary transformations that i) prepare the initial state $\Ket{\psi_0}$, ii) drive the desired evolution through discrete time steps $U(\delta t)=\mathrm{exp}(-i H \delta t)$, which are repeated $k$ times to simulate time evolution from $t=0$ to $t=k\delta t$, and iii) map the observable POVM outcomes, here one-dimensional orthogonal projectors $\Pi_a=\KetBra{\phi_a}{\phi_a}$, onto the logical basis states $\Ket{F,m}$. After this last step, populations are measured. This provides good estimates of the probabilities $p_a=\trace{\rho \Pi_a}$, as well as the expectation value $\langle A \rangle = \sum_a p_a a$. Simulations are performed in parallel on $\sim 10^7$ atoms, giving excellent measurement statistics for probability distributions in any arbitrary basis. Further details about this quantum simulation platform are given in Appendix \ref{app:experimental}.\\ 

Even though our findings are largely independent of the details of the model Hamiltonian that is being simulated, here we focus on a particular many-body quantum system, the Lipkin-Meshkov-Glick (LMG) model \cite{lipkin1965} (see Appendix \ref{app:other_systems} for other cases). The LMG Hamiltonian describes the dynamics of a system of $N$ spin-$\frac{1}{2}$ particles with Ising-like interactions in a completely connected graph, and reads
\begin{equation}
H_{LMG}(s) = -\frac{B}{2} \sum\limits_{i=1}^N \sigma_z^{(i)}- \frac{\Lambda}{4N}\sum\limits_{i,j=1}^N \sigma_x^{(i)}\sigma_x^{(j)}.
\label{ec:lmg_hami}
\end{equation} 
\noindent This model has been extensively analyzed in the literature and is a paradigmatic example of a quantum system presenting both ground state and excited state phase transitions in the thermodynamic limit \cite{ribeiro2007,santos2016}. Recalling the collective spin operators $S_\alpha=\frac{1}{2}\sum_i \sigma_\alpha^{(i)}$ introduced previously, the LMG Hamiltonian can be written in more compact form as $H_{LMG}=-B\:S_z - (\Lambda/N)S_x^2$. Due to conservation of the total spin $S^2$, we can focus on the evolution within the subspace of maximum spin $S=N/2$, which is composed of states that are completely symmetric under particle exchange and has dimension $N+1$. In our experimental simulations, we use $N=15$ to make use of the maximum Hilbert space size available with our platform. \\

\begin{figure*}[t!]
	\centering
	\includegraphics[width=0.95\linewidth]{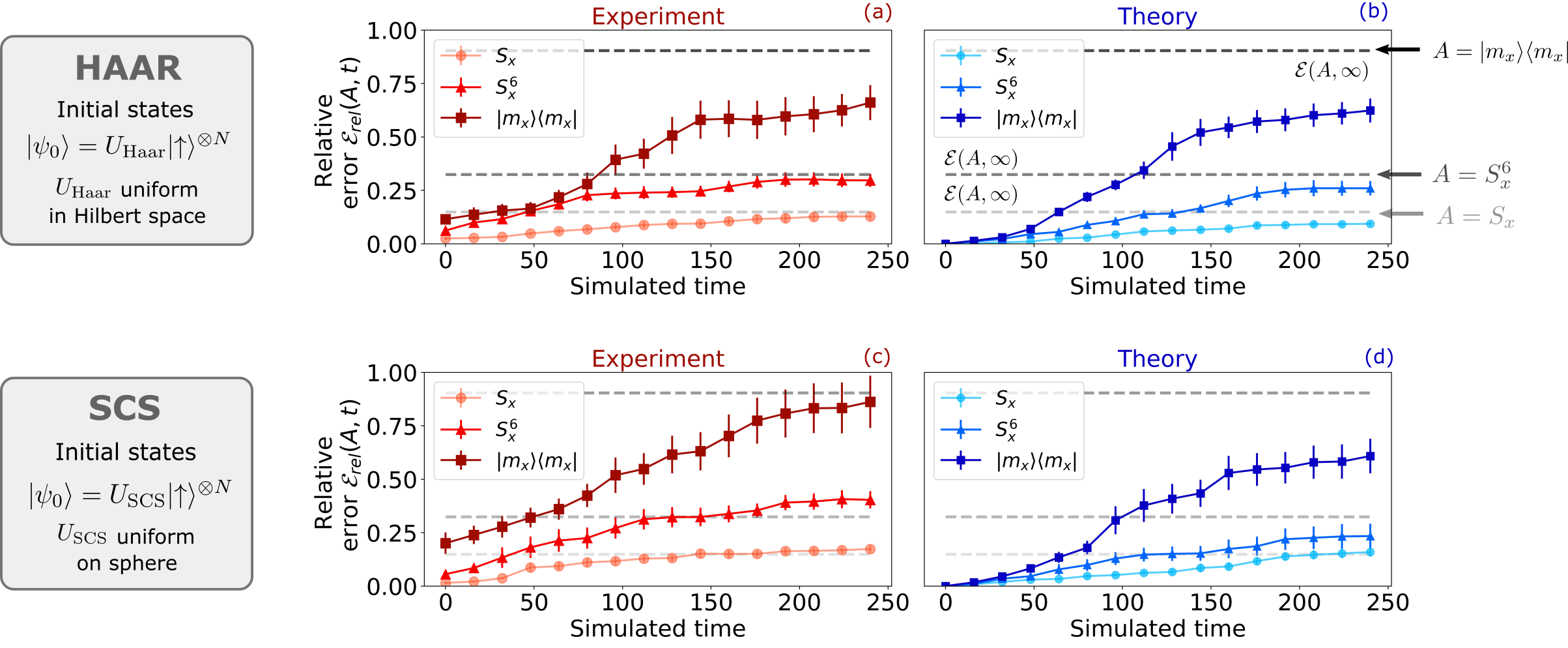}
	\caption{{\footnotesize Cumulative relative error in expectation values, corresponding to: \textbf{(a)}, \textbf{(c)} experimental quantum simulation results and \textbf{(b)}, \textbf{(d)} numerical simulations from the theoretical model. Results shown are averages over 10 initial states, chosen randomly from \textbf{(a)-(b)} the uniform (Haar) distribution over Hilbert space and \textbf{(c)-(d)} the uniform distribution of (separable) spin coherent states. In all cases, results for three different output observables are portrayed, which are chosen to have increasing purity: $A=$ $S_x$, $S_x^6$ and $\KetBra{m_x}{m_x}$ (here $m_x=\frac{1}{2}$, although similar results are obtained for other cases). Dashed lines correspond to the theoretical prediction for the asymptotic value of the error $\mathcal{E}^\infty_{rel}(A)$ for Haar-random initial states, c.f. Eq. (\ref{ec:err_inf}), computed for each $A$. Note that the same dashed lines are also included in the spin coherent states plots, as a guide to the eye.
    Error bars show standard error of the mean, arising from the averaging procedure over $n_s=10$ random initial states. For each initial state, the results corresponding to the theoretical model are obtained numerically using the same procedure described for the results in Fig. \ref{fig:Fig3}. All cases shown correspond to simulation of the LMG evolution, for $B=0.4\Lambda$. Simulated time has units of $1/\Lambda$.}}
	\label{fig:Fig4}
\end{figure*}

In Fig. \ref{fig:Fig3} we show the evolution of the fidelity and expectation values obtained from quantum simulations (in red) of the LMG dynamics for $B=1.5\Lambda$ (paramagnetic) and $B=0.4\Lambda$ (ferromagnetic), starting from a spin coherent state $\Ket{\psi_0}=\Ket{\downarrow_x}^{\otimes N}$. In Fig. \ref{fig:Fig3} (b) and (e) we plot the experimental infidelity, which reaches roughly $\sim 30-40\%$ at time $\Lambda t=30$, indicating significant deviation from the ideal quantum evolution. Nonetheless, expectation values are tracked with high accuracy, as can be seen from subplots (c) and (f) and the corresponding time-dependent errors in (d) and (g), which are seen to fluctuate over time (additional cases are presented in Appendix \ref{app:exp_extra}). This is the behavior predicted by our theoretical framework.\\

For comparison, we show results obtained from numerical simulations of the LMG Hamiltonian, combined with the theoretical model introduced in the previous section to include the effect of imperfections in the simulation (in blue). To obtain this, starting from the initial state $\Ket{\psi_0}$, the system is evolved with a perturbed Hamiltonian $H_{LMG}(s)+\lambda\:V$, where $V$ is taken as a random real matrix from the Gaussian Orthogonal Ensemble (GOE). Results shown correspond to averages over 50 instances of the random perturbation. The only free parameter is the perturbation strength $\lambda$ which is chosen to fit the infidelity curve (details about this procedure are presented in Appendix \ref{app:experimental}). Once $\lambda$ is fixed, the model reproduces the main features of the expectation value error curves as can be seen from Fig. \ref{fig:Fig3}(d) and (g). Notice that this is achieved without any information about the actual physical origin of the imperfections affecting the experiment, since the model only assumes that the diagonal matrix elements of the perturbing Hamiltonian in the basis of the ideal Hamiltonian are randomly distributed. The agreement between the experiment and the theoretical prediction is particularly striking, as the actual perturbations in the laboratory are expected to have fundamentally different character than the full random matrix perturbations used in our numerical simulations.\\

In order to explore the impact of imperfections on different observables $A$ as a function of their purity $\eta(A)$, we study the long-time average cumulative error associated with the experimental expectation value curves. Here, evolution times were taken to be about ten times larger than those shown in Fig. \ref{fig:Fig3}, in order to be closer to the asymptotic regime and enable comparison with our theoretical findings, c.f. Eq. (\ref{ec:err_inf}). To remove the dependence on the initial state of the system, we have obtained results for several different initial states, which we divide in two groups. First, a set $n_s=10$ random states sampled uniformly over the Haar measure, enabling direct comparison with the analytical result of Eq. (\ref{ec:err_inf}). Conceptually, however, it could be argued that these states are not physically relevant for AQS \cite{poulin2011}. We thus consider also a set of $n_s=10$ states which are prepared as (uniformly distributed) random rotations from the fiducial state $\Ket{\uparrow_x}^{\otimes N}$, leading to a set of random spin coherent states, which are typical initial conditions for AQS in the LMG model.\\

In Fig. \ref{fig:Fig4} we show the cumulative relative error, defined in Eq. (\ref{ec:error_rel_At}), averaged over Haar-random states in (a)-(b), and over spin coherent states in (c)-(d). Errors calculated from the experimental quantum simulations are shown in shades of red in (a) and (c), while those obtained from numerical simulations based on our random perturbation theoretical model are shown in shades of blue in (b) and (d). In all cases, we display results for three output observables: $A_1=S_x$, $A_2=S_x^6$ and $A_3=\KetBra{m_x}{m_x}$ (here $m_x=1/2$; other cases shown in Appendix \ref{app:exp_extra}). These observables are chosen to have increasing purity, as discussed in the beginning of this paper. For both sets of initial states, the errors generated in the experiment follow essentially the same behavior as the theoretical prediction.\\

For short times, when the cumulative errors are small, the experimental values deviate considerably from the numerical curves. Most of this difference can be attributed to state preparation and measurement (SPAM) errors, as we show in Appendix \ref{app:spam}. Nevertheless, for longer times the relative errors in the expectation values become consistently higher as the purity of the corresponding observable increases. This confirms the role of the purity as a measure of sensitivity of expectation values to imperfections in the state. For Haar-random initial states, the dashed lines indicate the analytical result of Eq. (\ref{ec:err_inf}), which is able to faithfully reproduce the asymptotic values of the cumulative relative error for all cases. In Appendices \ref{app:exp_extra}, \ref{app:num_extra} we present further experimental and numerical data that illustrates this behavior. Also, in Appendix \ref{app:other_systems} we present numerical simulations showing that our theoretical framework is generally applicable to other systems, like the transverse Ising model. \\

To further demonstrate the role that the observable purity plays in the buildup of errors, in Fig. \ref{fig:Fig5} we plot the long-time cumulative relative error computed for several choices ($\sim 10$) of output observable $A=S_x^{2k}$. As seen in Fig. \ref{fig:Fig1} (b), the purity of these observables increases monotonically with $k$. In Fig. \ref{fig:Fig4} we plot the long-time relative error as a function of the observable purity for (a) Haar-random initial states and (b) random spin coherent initial states. In both cases, it is evident that the errors are a monotonic function of the purity, which determines how well expectation values can be tracked in the presence of imperfections. Matching between experiment and theory is also observed, especially for the Haar-random initial states. In Fig. \ref{fig:Fig5} (c) we plot the same data as in (a) but as a function of a modified purity $\trace{\rho_A^2}-\trace{\rho_{A_D}^2}$, where we have computed the purity of the Hamiltonian-dependent observable $A_D$ from Eq. (\ref{ec:err_inf}). The resulting data can then be directly compared with the analytical prediction of Eq. (\ref{ec:err_inf}), showing very good agreement. \\

\section{Discussion}

\begin{figure}[t!]
	\centering
	\includegraphics[width=0.95\linewidth]{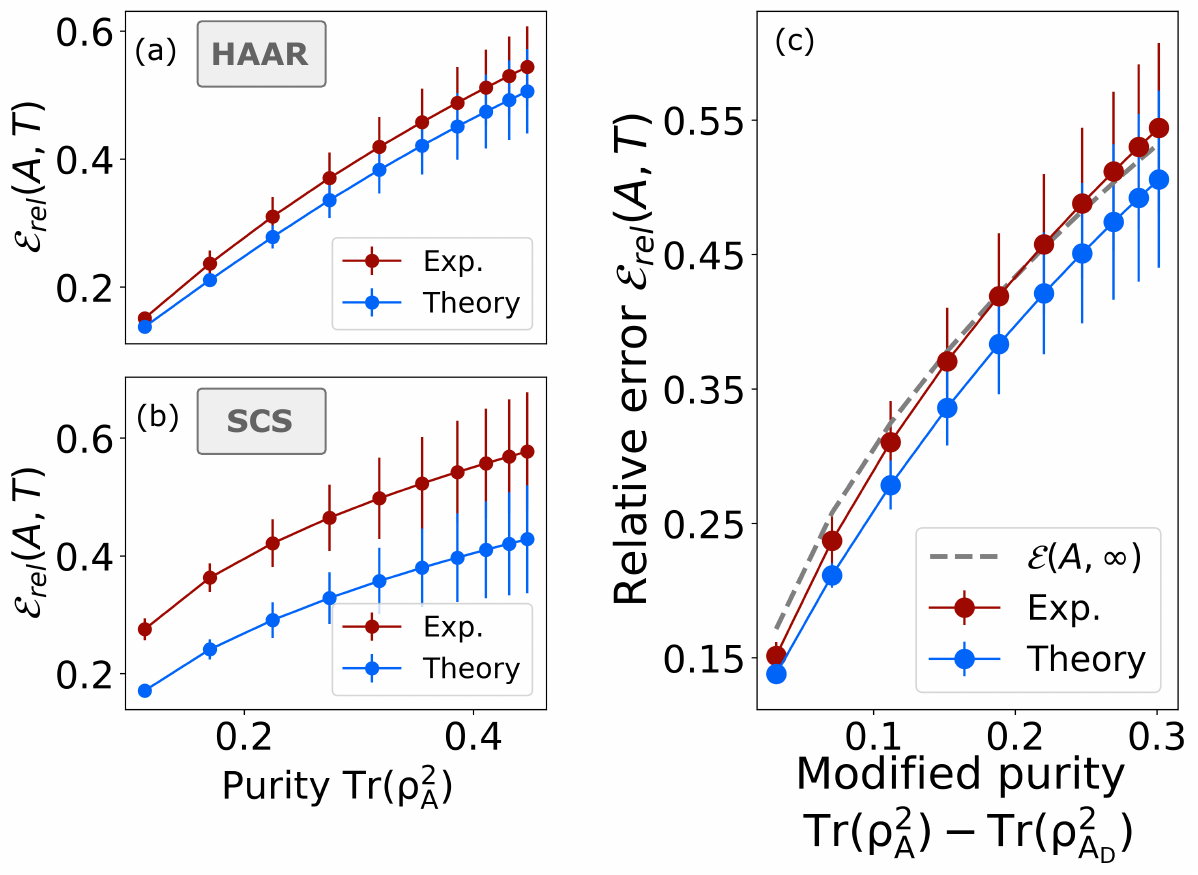}
	\caption{{\footnotesize \textbf{(a)} - \textbf{(b)} Average relative error as a function of purity of the output observable. In all cases, points correspond to the long-time value of the cumulative relative error, c.f. Eq. (\ref{ec:rms_error}), computed from experimental quantum simulation data (in red) and numerical simulations from the theoretical model (in blue). Cases portrayed correspond to \textbf{(a)} Haar-random initial states and \textbf{(b)} random spin coherent initial states. \textbf{(c)} Same data as in \textbf{(a)} but plotted against the Hamiltonian-dependent modified purity $\trace{\rho_A^2} - \trace{\rho_{A_D}^2}$, to compare with the theoretical prediction for the asymptotic error of Eq. (\ref{ec:err_inf}) (gray dashed line). All data shown corresponds to simulations of the LMG Hamiltonian with $B=0.4\Lambda$. Error bars denote standard error of the mean as in Fig. \ref{fig:Fig4}.}}
	\label{fig:Fig5}
\end{figure}

In this work we have presented a framework to characterize the effect of errors in the output of quantum simulators. Critically, we have introduced the observable purity $\eta(A)$ as the key metric that can be used to characterize the average sensitivity of expectation values $\langle A\rangle$ to random imperfections. By performing extensive experimental explorations in a small-scale universal quantum simulator, we were able to demonstrate the validity of this framework in a real-world device, without using any knowledge about the nature of the imperfections affecting its operation.\\ 

Beyond describing the dependence of errors on the output observable, we argue that the presented framework can be regarded as a tool to predict the expected robustness of different aspects of quantum simulations. For instance, the observable purity $\eta(A)$, which determines the long time error, c.f. Eq. (\ref{ec:err_inf}), is expected to be a known quantity even in instances where the simulation is classically intractable (if its spectrum is known, as it is expected for most physical observables). Conversely, the short-time behavior of the simulation error is also shown to be properly captured by the proposed model. Thus, once the perturbation parameter $\lambda$ is found experimentally, it becomes possible to make quantitative predictions about various aspects of the simulation.  For example, one can estimate the time frame within which the simulation of the selected observable can be trusted to fit within a given error budget. Also, one can model the tradeoff between the native errors of the device and those that arise from other sources, for instance actual approximations made in the programming of the device (such as Trotter errors). Most importantly, this can be achieved without having to do the hard work of unraveling the actual, physical errors that happen on the specific hardware.\\

While we have tested our predictions on a specific quantum information processing platform, we expect our framework to be broadly applicable, as the concept of typicality \cite{bartsch2009,reimann2019} describes intrinsic robustness of expectation values for a wide range of perturbations. In scenarios with specific kinds of imperfections, e.g. local decoherence on a simulator with a local tensor product structure, errors will also depend on additional properties of the output observable. For some instances, given specifically chosen initial states of the simulator, we expect to see deviations from the universal average behavior set by $\eta(A)$. Nonetheless, we expect that the observable purity is the central tool to establish the baseline sensitivity of AQS outputs. More generally, we also anticipate that this framework will have applications in topics like tomography and sampling \cite{paini2019,heger2019,huang2020} and relaxation in many-body systems \cite{dabelow2020}.

\section*{Acknowledgments}

PMP acknowledges Diego Wisniacki for insightful discussions. This work was supported by the U.S. National Science
Foundation Grants No. 1630114, No. 1521439, No. 1820679, and No. 1820758.\\


%
%

\appendix
\section{State-averaged simulator error} \label{app:static}

In the case that the simulated state $\Ket{\psi_{\rm sim}}$ is pure, the simulator error of Eq. (\ref{ec:errorA}) can be written as
\begin{equation}
\delta(A)=\mathcal{N}(\gamma)^2\left[\gamma^2\left(\langle A\rangle -\langle A\rangle_\perp\right)-2\gamma \mathrm{Re}\left(\Bra{\psi_\perp} A\Ket{\psi}\right)\right]
\label{ec:met_errorA}
\end{equation}

\noindent We now write the states above as $\Ket{\psi}=U\Ket{a_l}$ and $\Ket{\psi_\perp}=U\Ket{a_m}$, where $l\neq m$ and $\left\{\Ket{a_i}\right\}$ with $i=1,\ldots,d$ is a reference basis set (for instance, the basis of eigenstates of $A$). Here, $U$ is a random matrix taken from the uniform distribution over the manifold of $d\times d$ unitary matrices, i.e. the Haar measure corresponding to the group $\mathrm{U}(d)$ \cite{hall2015}. We can then consider quantities like
\begin{equation}
\int_{\mathrm{U}(d)} \delta(A)^k \: dU \equiv \overline{\delta(A)^k},
\end{equation}

\noindent which can be computed analytically (for $k=1,2$) decomposing the integrand into sums of polynomials in the elements of $U$. Using known expressions for these integrals for up to quartic order \cite{collins2006}, which we show in Appendix \ref{app:purity}, we can compute
\begin{eqnarray}
\overline{\langle A\rangle}  &=& \frac{1}{d}\trace{A} \label{ec:avgA1} \\
\overline{\langle A\rangle^2} &=& \frac{1}{d^2+d}\left(\trace{A^2}+\trace{A}^2\right) \label{ec:avgA2}\\
\overline{\langle A\rangle \langle A\rangle_\perp} &=& \frac{1}{d^2-1}\left(\trace{A}^2-\frac{1}{d}\trace{A^2}\right) \label{ec:avgA3} \\
\overline{|\Bra{\psi} A\Ket{\psi_\perp}|^2} &=& \frac{1}{d^2-1}\left(\trace{A^2}-\frac{1}{d}\trace{A}^2\right).  \label{ec:avgA4}
\end{eqnarray}

\noindent Inserting these results in Eq. (\ref{ec:met_errorA}) we obtain $\overline{\delta(A)}=0$ and the expression for $\overline{\delta(A)^2}$ in  Eq. (\ref{ec:errorA_avg}).

Furthermore, these allow us to easily generalize our findings to mixed states, i.e. $\Ket{\psi_{\rm sim}}\rightarrow \rho_{\rm sim}$. For one simple noise model
\begin{eqnarray}
\rho_{\rm sim} = (1-\gamma)\KetBra{\psi}{\psi} + \frac{\gamma}{d}\mathbb{I},
\end{eqnarray}

\noindent which has been discussed recently in the context of near-term quantum information processing devices \cite{arute2019}, we obtain  $\delta(A)=\langle A\rangle - \trace{\rho_{\rm sim} A}$ which evaluates to $\delta(A)=\gamma\left( \langle A\rangle - \overline{\langle A\rangle}\right)$. Using Eqs. (\ref{ec:avgA1})-(\ref{ec:avgA4}), it is straightforward to derive the average relative error for this model
\begin{equation}
\delta_{rel}(A)=\sqrt{\frac{d}{d+1}\gamma^2 \left(\trace{\rho_A^2}-\frac{1}{d}\right)},
\end{equation}

\noindent which depends on the purity in the same way as Eq. (\ref{ec:errorA_avg_rel}). A very similar result can be derived for the alternative the model $\rho_{\rm sim} = (1-\gamma)\KetBra{\psi}{\psi} + \gamma\KetBra{\psi_\perp}{\psi_\perp}$.\\

\section{Perturbation model for dynamical quantum simulators} \label{app:dynamic}

The time-dependent simulator error Eq. (\ref{ec:errorAt_def}) can be written in terms of the quantum state averaged over the perturbation, 
\begin{equation}
\left[ \KetBra{\psi_{\rm sim}(t)}{\psi_{\rm sim}(t)}\right]_V \equiv [\rho_{\rm sim}(t)]_V = \left[ U'(t) \rho_0 U'^\dagger(t)\right]_V
\end{equation}

\noindent where we have denoted $U'(t)=\ex{-i(H + \lambda V)t}$ and $\rho_0=\KetBra{\psi_0}{\psi_0}$. Using time-independent perturbation theory to expand the eigenvalues and eigenvectors of $H'=H +\lambda V$ in powers of $\lambda$, we obtain
\begin{equation}
U'(t) = U_V(t) + \mathcal{O}(\lambda)
\end{equation}

\noindent where $U_V(t)$ is a unitary matrix given by
\begin{equation}
U_V(t)=\sum\limits_k \ex{-i(E_k+\lambda V_{kk}) t}\KetBra{u_k}{u_k}.
\label{ec:met_Uv}
\end{equation}

Notice that the zeroth order contribution depends on the perturbation through its diagonal elements $V_{kk}$. This is because $\lambda$ cannot be neglected in the argument of the exponential since the time $t$ could in principle be of order $\lambda^{-1}$ \cite{peres1984}. Thus, the leading order contribution to the perturbed state is
\begin{equation}
\left[ \rho_{\rm sim}(t)\right]_V = \left[U_V(t) \rho_0 U_V(t)^\dagger\right]_V + \mathcal{O}(\lambda).
\label{ec:met_rhosim}
\end{equation}

From Eqs. (\ref{ec:met_Uv}) and (\ref{ec:met_rhosim}) it can be seen that the effect of the perturbation is condensed on the quantity 
\begin{equation}
f(\tau) = \left[ \ex{-i(V_{ll} - V_{mm})\lambda t} \right]_V = \left| \int  p_V(x) \ex{-i \tau x} dv \right|^2 = \left| g(\tau)\right|^2
\label{ec:ftau_pert}
\end{equation}

\noindent where $l\neq m$, we have defined $\tau\equiv \lambda t$ and introduced $p_V(x)$ which is the probability density function associated with the perturbation matrix elements $V_{kk}$. A crucial assumption here is that the diagonal matrix elements $V_{kk}$ can be considered statistically independent. The integral inside the absolute value in Eq. (\ref{ec:ftau_pert}) is the \textit{characteristic function} $g(\tau)$ of the probability distribution \cite{kropf2016}, which has the properties $g(0)=1$ and $\lvert g(\tau)\rvert \leq 1$ \cite{loeve1963}. Furthermore, for all probability distributions of interest, $f(\tau)$ will be a function that goes to 0 as $\tau\rightarrow \infty$. As an example, if $V$ is taken to be a random matrix from the Gaussian orthogonal ensemble (GOE), then we have that $g(\tau)=\ex{-\tau^2/2}$ \cite{cerruti2003}. \\

After defining $f(\tau)$, we obtain for the perturbed state the following expression
\begin{equation}
\left[ \rho_{\rm sim}(t)\right]_V = \rho_{\psi,D} + f(\tau)\left(\rho(t)-\rho_{\psi,D}\right),
\end{equation}

\noindent from which the result on Eq. (\ref{ec:errorAt}) immediately follows.

\section{Simulation on a quantum processor} \label{app:experimental}

\noindent \textit{Quantum control.} As described in the main text, our SHAQ simulator is based on the spin degrees of freedom of an individual $^{133}$Cs atom in the electronic ground state. The atom is driven by a combination of static and time varying magnetic fields, rendering it fully controllable in a 16-dimensional Hilbert space. Control is achieved with a combination of a static magnetic bias field along $z$, a pair of phase modulated radio-frequency (rf) magnetic fields along $x$ and $y$, and a single phase-modulated microwave ($\mu$w) magnetic field.  The rf fields are tuned to the Larmor precession frequency in the bias field, and the microwave field is tuned to the transition between the $\Ket{3,3}$ and $\Ket{4,4}$ states. Phase-modulation waveforms that implement a desired transformation $U\in \mathrm{SU}(16)$ are found using either conventional optimal control, or a variant optimized for AQS that searches for co-optimal control fields and system-simulator maps using a new approach called EigenValue-Only (EVO) control \cite{lysne2020}. The two protocols generate (non-unique) control waveforms consisting of 150 and 20 discrete phase steps, respectively, corresponding to waveform durations of 600$\mu$s and 80$\mu$s, with typical fidelities $\mathcal{F}=0.985$ and $\mathcal{F}=0.995$ as estimated by randomized benchmarking.  EVO control is used exclusively for the unitary time steps that make up an AQS, while conventional control is used to generate unitary maps for state preparation and measurement.\\

\noindent \textit{Experimental Implementation.} Our experiment begins with a sample of ${\sim} 10^7$ laser cooled Cs atoms released from a magneto-optical trap/optical molasses into free fall. Optical pumping and state selective purification is used to prepare $>99\%$ of these atoms in the logical basis state $\Ket{\chi_0}=\Ket{3,3}$. An AQS sequence begins with a map $W_{\rm SP}$ to the desired input state, $\Ket{\chi_0}\rightarrow \Ket{\psi_0}=\sum_{F,m} C_{F,m}\Ket{F,m}$. This is followed by $k$ identical time steps $U$ of duration $\delta t$ to simulate time evolution from $t=0$ to $t=k\delta t$. Finally, to measure a desired observable $A=\sum_a a \KetBra{\phi_a}{\phi_a}$ we apply a unitary map $W_\mathrm{M}=\sum_a \KetBra{(F,m)_a}{\phi_a}$, and determine the population of the states $\Ket{(F,m)_a}$ with a Stern-Gerlach measurement.  The latter is implemented by imposing a magnetic field gradient on the falling atoms and measuring the state dependent arrival times at a resonant probe beam located below the preparation volume. Fitting the time dependent fluorescence from atoms falling through the probe gives an accurate measure of the populations in the logical basis states $\Ket{F,m}$, and these in turn provide good estimates of the probabilities $p_a=\trace{\rho \Pi_a}$ for the POVM outcomes $\Pi_a=\KetBra{\phi_a}{\phi_a}$, and the expectation value $\langle A \rangle = \sum_a a p_a$. Measurement statistics contribute negligibly due to the large number of simulations running in parallel on millions of atoms; instead the accuracy is dominated by errors in the readout map, probe power fluctuations, and electronic noise.  Measuring the projector $\KetBra{\chi_k}{\chi_k}$, where $\Ket{\chi_k}$ is the state predicted in the absence of errors, gives an estimate of the fidelity of the AQS. Averaged over a sample of random states, the SPAM error on this estimate is approximately 1\%.  For a detailed evaluation of SPAM errors, see Appendix \ref{app:spam}. For additional information about the operation and performance of our SHAQ simulator, see \cite{lysne2020}.\\

\noindent \textit{Estimating the perturbation strength $\lambda$.} When $V$ is a random matrix taken from the GOE, we can use the explicit form for $f(t)$ with Eq. (\ref{ec:infid}) to get
\begin{equation}
\delta (\rho(t),t) = \mathcal{I}(t) = (1 - e^{-\tau^2} ) ( 1 - S_{0}).
\label{ec:infidGOE}
\end{equation}
In this case, the growth in infidelity of the model depends directly on the perturbation strength $\lambda$, as $\tau = \lambda t$ and $S_{0}$ is fixed by the choice of initial state. This gives us a way to fix $\lambda$ based on the decay of the state-level fidelity in the experiment, which shares the same form to leading order \cite{knill2008}. We use the ability to perform arbitrary unitary transformations to map the state $\Ket{\psi(t)}$ to a logical basis state. The resulting measurement is of the observable  $A=\KetBra{\psi(t)}{\psi(t)}$, from which we can calculate the infidelity. We then fit the data to minimize the residual between the data and Eq. (\ref{ec:infidGOE}) to find the $\lambda$ that most closely matches the growth in experimental infidelity. Typical fit values are around $\lambda \approx 0.01$, suggesting the amount of perturbation from the ideal is small.


\section{Analysis of state preparation and measurement (SPAM) errors} \label{app:spam}


\begin{figure}[t]
	\centering
	\includegraphics[width=1.\linewidth]{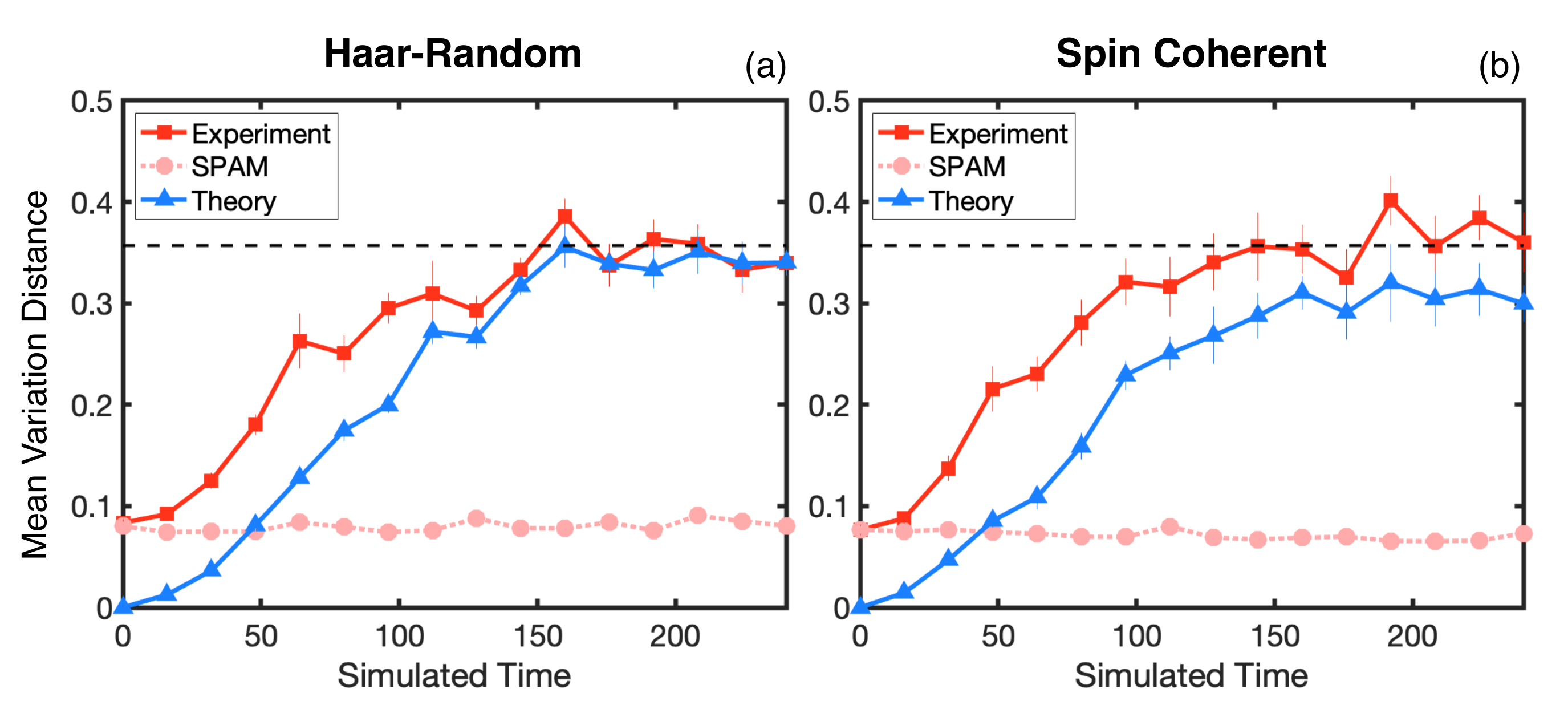}
	\caption{{\footnotesize Mean Variation Distance to the exact probability distributions calculated over sets of $n_s=10$ (a) Haar-random and (b) random spin coherent initial states. Squares indicate distances for the experimental quantum simulation.  Circles indicate distances for experimentally obtained SPAM errors. Triangles indicate distances for the theoretical simulations.  All error bars represent the standard error of the mean.  The black dashed line is a numerically calculated mean distance of the mixed state population distribution from $10^4$ Haar-random states.}}
	\label{fig:spam1}
\end{figure}


When running an analog quantum simulation on a physical device, errors are introduced through imperfections in the evolution (simulation errors) and through imperfect state preparation and measurement (SPAM errors). The theoretical model developed in the main text considers only simulation errors, whereas our experiment is subject to both simulation and SPAM errors. In this section we examine SPAM errors and the role they play in our experiment. 

While SPAM errors cannot be eliminated, their importance can be evaluated by comparing experiments with and without simulation errors.  To see how, consider that an experimental quantum simulation has three parts: (i) preparation of a chosen input state, (ii) an iterative sequence of $k$ time steps, and (iii) an orthogonal measurement at the end.  Thus, starting from a given input state, we can calculate the output state resulting from an ideal simulation, forego part (ii) and instead prepare this state directly, and proceed with the measurement.  The resulting error is then a good measure for the SPAM error in a $k$-step simulation.  


Fundamentally, the output of a quantum simulator is a probability distribution over measurement outcomes. In this case it is appropriate to use the total variation distance $D_V$ as a metric for error. For two probability distributions $P:\{p_n\}$ and $Q:\{q_n\}$, the variation distance is given by
\begin{equation}
D_V = \frac{1}{2}\sum\limits_{n=1}^d |p_n - q_n|
\label{ec:tot_var_dist}
\end{equation}

\noindent where $d=16$ is the dimension of the Hilbert space. In Fig. \ref{fig:spam1} we show the total variation distance between the probability distributions calculated for an ideal simulation and those of the experiment (which include SPAM and simulation errors), SPAM error only, and the predictions of our theory (only affected by simulation errors) as a function of simulated time. As in Fig. 3 in the main text, results are averaged over $n_s=10$ initial states. We see that at short times, variation distances for the experiment is almost entirely due to SPAM errors, in the case of Haar random as well as spin coherent initial states. At late times, simulations with Haar-random initial states saturate the total error in both the experiment and the theory, irrespective of the presence or absence of SPAM errors, suggesting that the two types of error become uncorrelated as simulation errors accrue. For simulations with spin coherent initial states the errors also approach saturation, but experiment and theory do not appear to converge with in the time simulated. 

These behaviors are also manifest in the cumulative relative errors of the observables examined in the main text, namely $A=S_x$, $A=S_x^6$ and $A=\KetBra{m_x=\frac{1}{2}}{m_x=\frac{1}{2}}$, and can be seen in Figure \ref{fig:spam2}.  It is important to note that both the mean variation distances and the observable relative errors are calculated using the same probability distributions, and that the two are therefore likely to correlate. Furthermore, one can expect the obervable error to depend on observable purity in roughly the same way, regardless of whether the underlying cause is simulation or spam error.  Both trends are clearly supported by the data. 

\begin{figure*}[t]
	\centering
	\includegraphics[width=0.8\linewidth]{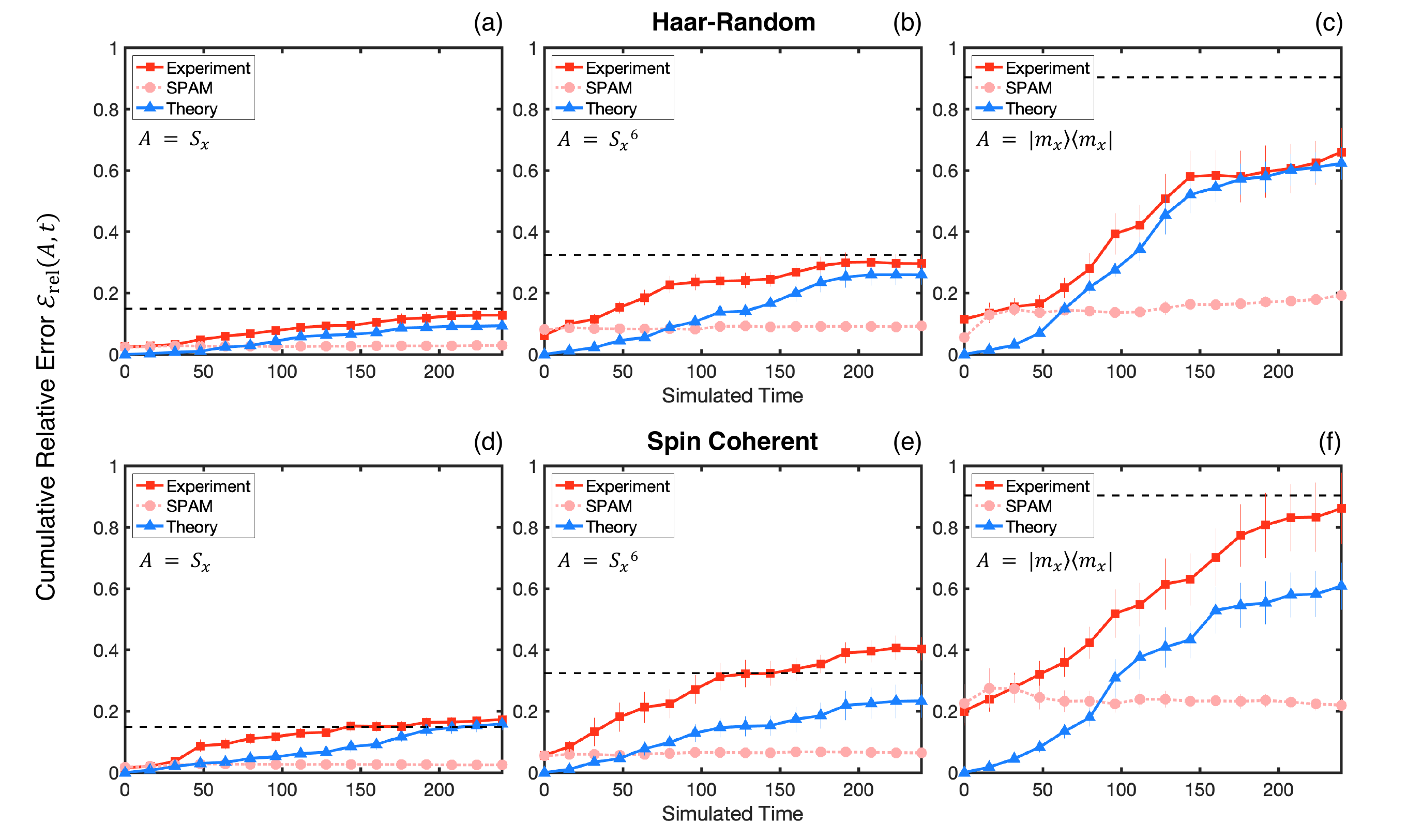}
	\caption{{\footnotesize Cumulative relative error over sets of $n_s=10$ Haar-random (a-c) and spin coherent (d-f) initial states.  Squares indicate deviations for the experimental quantum simulation.  Circles indicate results for experimentally estimated SPAM errors. Triangles indicate errors for the theoretical-numerical simulations.  All error bars represent the standard error of the mean.  The black dashed lines are the theoretical predictions for the asymptotic value of the error $\mathcal{E}_{\mathrm{rel}(A,\infty)}$ for Haar-random initial states, c.f. Eq. (\ref{ec:err_inf}).}}
	\label{fig:spam2}
\end{figure*}

\section{Observable purity analysis} \label{app:purity}

\subsection{Average and variance analysis}

In this section we analyze particular aspects of the simulator error $\delta(A)$, defined in Eq. (1) of the main text, when evaluated for Haar-random states. Formally, we are interested in quantities like 
\begin{equation}
\int_{\mathrm{U}(d)} \delta(A)^k \: dU \equiv \overline{\delta(A)^k},
\label{ec:sm_int_haar}
\end{equation}

It is easy to see that Eq. (\ref{ec:sm_int_haar}) can be decomposed into linear combinations involving the $k$-th order moments of the Haar distribution \cite{collins2006,puchala2017}
\begin{equation}
\mathcal{P}_k = \int_{\mathrm{U}(d)}  U_{i_1 j_1}\ldots U_{i_k j_k}U_{i_1' j_1'}^*\ldots U_{i_k' j_k'}^*\: dU.
\label{ec:sm_haar_moments}
\end{equation}

In order to derive Eqs. (2)-(3) and (10) in the main text we have used the expressions for $k=1$ and $k=2$, which are relatively easy to obtain in closed form and read \cite{roberts2017}
\begin{eqnarray}
\mathcal{P}_1 &=& \frac{1}{d}\delta_{i_1 i_1'}\delta_{j_1 j_1'}\\
\mathcal{P}_2 &=& \frac{1}{d^2-1}\left(\delta_{i_1 i_1'} \delta_{i_2 i_2'}\delta_{j_1 j_1'}\delta_{j_2 j_2'} + \delta_{i_1 i_2'} \delta_{i_2 i_1'}\delta_{j_1 j_2'}\delta_{j_2 j_1'}\right) - \\
& &\frac{1}{d(d^2-1)}\left(\delta_{i_1 i_1'} \delta_{i_2 i_2'}\delta_{j_1 j_2'}\delta_{j_2 j_1'} + \delta_{i_1 i_2'} \delta_{i_2 i_1'}\delta_{j_1 j_1'}\delta_{j_2 j_2'}\right).
\end{eqnarray}

These calculations lead to the main result for the average relative error,
\begin{equation}
\delta_{rel}(A)^2 \equiv \frac{\overline{\delta(A)^2}}{\overline{\langle A \rangle}^2} =  \frac{2d^2}{d^2-1}\left(\frac{\gamma^2}{1+\gamma^2}\right)\left(\trace{\rho_A^2}-\frac{1}{d}\right),
\label{ec:sm_err_rel}
\end{equation}

\noindent c.f. Eq. (3) in the main text. To numerically illustrate this result, we consider observables corresponding to a system of $N$ spin-$\frac{1}{2}$ particles, like the collective magnetization $S_z$. Restricted to the symmetric subspace, the Hilbert space dimension for this case is $d=N+1$. Then, we evaluate the relative error in the expectation value of different observables for a set of 2000 Haar-random states (with corresponding random perturbations). The numerically computed average is shown in Fig. \ref{fig:sm_haar_var} (a), together with the analytical result of Eq. (\ref{ec:sm_err_rel}). The set of observables is chosen as $S_z^\alpha$, with $\alpha$ even except for $\alpha=1$. From this plot we can see a monotonic behavior resembling $d^{-1}$ for large $N$ and the expected increasing relative error as the higher the power (and hence, the purity) of $S_z$ is considered (recall Fig. 1 in the main text).\\

\begin{figure}[t]
	\centering
	\includegraphics[width=1.\linewidth]{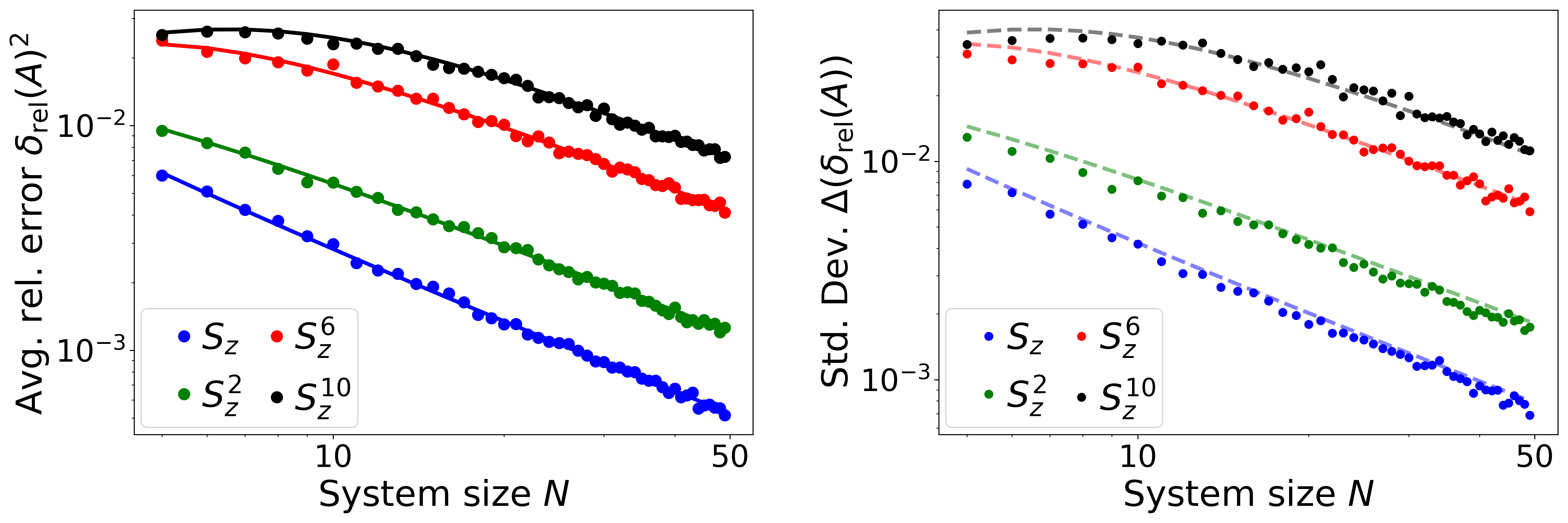}
	\caption{{\footnotesize \textbf{(a)} Average value of the relative error $\delta_{\rm rel}(A)^2$ calculated numerically over 2000 random states $\Ket{\psi}$ and $\Ket{\psi_{\rm sim}}=\mathcal{N}(\gamma)(\Ket{\psi} + \gamma \Ket{\psi_\perp})$. States $\Ket{\psi}$ and $\Ket{\psi_\perp}$ are drawn randomly from the Haar distribution, and here we take $\gamma=0.2$. Results are shown for four different observables, showing higher average relative errors as the observable purity is increased. Full lines denote the analytical result of Eq. (\ref{ec:sm_err_rel}). \textbf{(b)} Numerically calculated standard deviation of the quantity $\delta_{\rm rel}(A)^2$, showing the magnitude of fluctuations around the mean shown in \textbf{(a)}. The faded dashed lines represent correspond to $1.5\times\overline{\delta(A)^2}$ to illustrate the similarity of scaling between the standard deviation and the average value, particularly for large system size. }}
	\label{fig:sm_haar_var}
\end{figure}

An important part of this analysis is to assess how big the are the deviations from the Haar-average value, as a function of the observable. For this, we need consider the variance of $\delta(A)^2$, i.e.
\begin{equation}
\Delta[\delta(A)^2]^2\equiv \overline{\delta(A)^4} - \overline{\delta(A)^2}^2 .
\end{equation}

Notice that the explicit calculation of $\overline{\delta(A)^4}$ involves 4th order moments of the Haar distribution, c.f. Eq. (\ref{ec:sm_haar_moments}). Since it is tricky to obtain closed expressions for these, we rely on full numerical analysis. Results are shown in Fig. \ref{fig:sm_haar_var} (b), where we show $\Delta[\delta_{\rm rel}(A)^2]$ .  We observe that higher purity observables, for which the average relative error is higher, also have higher fluctuations, and conversely low purity observables, which have small average relative errors, have smaller fluctuations. In all cases, we observe that fluctuations decay with system size, suggesting that the average behavior of $\delta(A)$ is also typical for random states. Notably, the dependence of $\Delta[\delta_{\rm rel}(A)^2]$ on both the observable $A$ and the system size $N$ resembles very closely that of the average value. This implies that the ratio between the size of the fluctuations and the magnitude of the relative error is roughly the same for all cases. 

\subsection{Purity of many-body observables}

Here we analyze the concept of observable purity in the context of a many-body system. For concreteness, we consider systems of $N$ spin-$\frac{1}{2}$ particles. An important observation is that all operators of the form

\begin{equation}
A'_{(i_1,i_2,\ldots,i_k)}\equiv A'_{\vec{i}} = \sigma_z^{(i_1)}\sigma_z^{(i_2)}\ldots \sigma_z^{(i_k)},
\end{equation}

\noindent where $i_1\neq i_2 \neq \ldots i_k$ and $k\leq N$ denotes the number of sites the operator acts nontrivially on, have exactly the same spectrum. This can be easily seen by the fact that their eigenvalues are $-1$ and $1$, and that $\trace{A'_{\vec{i}}}=0$. This necessarily impies that half of the eigenvalues are $-1$ while the other half are $1$. Defining $A_{\vec{i}}=A'_{\vec{i}}+1$ (to make the operator positive), we can easily calculate the observable purity. Independently of $k$, all these operators display the same purity, i.e.
\begin{equation}
\mathrm{If}\ \rho_{\vec{i}}=\frac{A_{\vec{i}}}{\trace{A_{\vec{i}}}}\ \Rightarrow \trace{\rho_{\vec{i}}^2}=\frac{2}{d}=2^{-(N-1)}
\end{equation}

As a consequence, the purity is not related to the local character of the observable. However, as stated in the main text, there is an association between macroscopicity and purity. For many-body systems, this can be revealed in the following way. Consider a partition of an $N$-particle system into $\mathcal{S}_k$ (having $k$ particles) and $\mathcal{S}_{N-k}$ (including the remainder). We can define the observable 
\begin{equation}
O(k) = \KetBra{\phi_k}{\phi_k}\otimes \mathbb{I}_{N-k} 
\end{equation}
\noindent where $\Ket{\phi_k}$ is a pure state of $\mathcal{S}_k$ and $\mathbb{I}_{N-k}$ denotes the identity operator acting on $\mathcal{S}_{N-k}$. The purity for this observable is
\begin{equation}
\trace{\rho_{O}^2} = 2^{k-N}.
\end{equation}
From this expression we observe that when $k\sim  N$, the observable value sharply depends on the microstate of the system and $\trace{\rho_{O}^2}\sim 1$. On the other hand when $k \ll N$ we associate it with a macroscopic observable and indeed we have $\trace{\rho_{O}^2}\sim 2^{-N}$.

\section{Additional experimental results} \label{app:exp_extra}
\begin{figure*}[t!]
	\centering
	\includegraphics[width=0.7\linewidth]{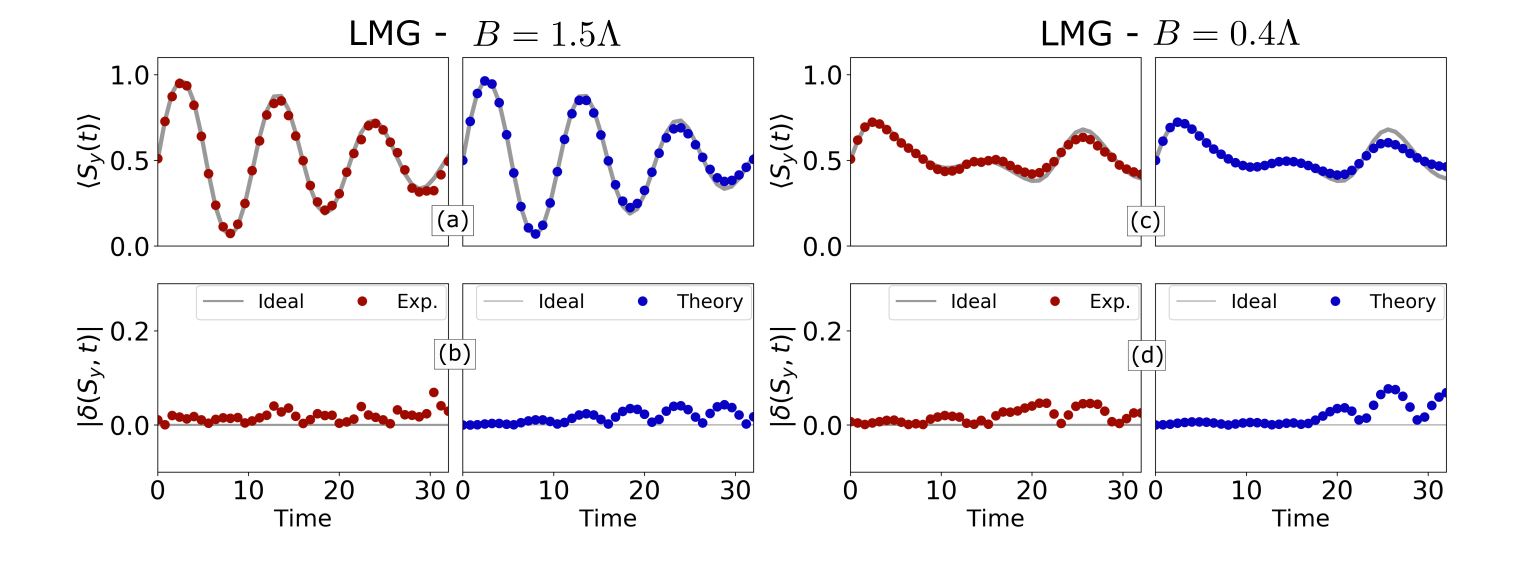}
	\caption{{\footnotesize Comparison between experimental quantum simulation results and the proposed model for dynamics of error generation (additional data corresponding to Fig. 3 of the main text). In all cases, red dots correspond to experimental results, and blue dots to numerical results obtained from the theoretical model, where the imperfection strength $\lambda$ is the only fitting parameter. Plots \textbf{(a)} and \textbf{(c)} show dynamics of expectation value of $S_y$ for experiment and theory. Full gray line is the ideal target evolution without errors. \textbf{(b)} and \textbf{(d)} show the time-dependent error in the expectation values $\delta(S_y,t)$ computed for these cases. Data in panel \textbf{(c)} is also shown in Ref. \cite{lysne2020}.}}
	\label{fig:fig_ext_fig2}
\end{figure*}
In this Section we present additional experimental data that complements the results shown in the main text. In Fig. \ref{fig:fig_ext_fig2} we show the evolution of $\langle S_y\rangle$ obtained from experimental simulations of the Lipkin-Meshkov-Glick (LMG) dynamics for parameter values $B=1.5\Lambda$ and $B=0.4\Lambda$. The initial state for both cases is $\Ket{\psi_0}=\Ket{\uparrow_x}^{\otimes N}$. On Fig. \ref{fig:fig_ext_fig2} (b) and (d) we plot the error, i.e. the difference between the simulated and the ideal outputs, for both experimental data and numerical results obtained from the theoretical model. As mentioned in the main text, we observe that the model reproduces the main features of the of the error dynamics.\\

In Fig. \ref{fig:fig3_extra} we show additional data corresponding to Fig. 4 of the main text. In particular, we plot the average relative error $\mathcal{E}_{rel}(P_m,t)$ evaluated for observables which are projectors onto pure states, $P_m=\KetBra{m_x}{m_x}$ for $m=-\frac{15}{2},-\frac{7}{2},\frac{7}{2},\frac{15}{2}$. Note that the case $m_x=\frac{1}{2}$ is shown in the main text. Results are presented for Haar-random initial states in \textbf{(a)-(b)} and for random spin coherent states in \textbf{(c)-(d)}. It can be seen from the figure that all these cases display a similar behavior and reach values which are comparable in magnitude, and considerably larger than the lower purity observables shown in Fig. 4 of the main text. This behavior agrees with the fact that all the observables $P_m$ have the same purity (equal to one), and thus their expectation values should display, on average, the same degree of sensitivity to imperfections.
\begin{figure*}
		\includegraphics[width=0.6\linewidth]{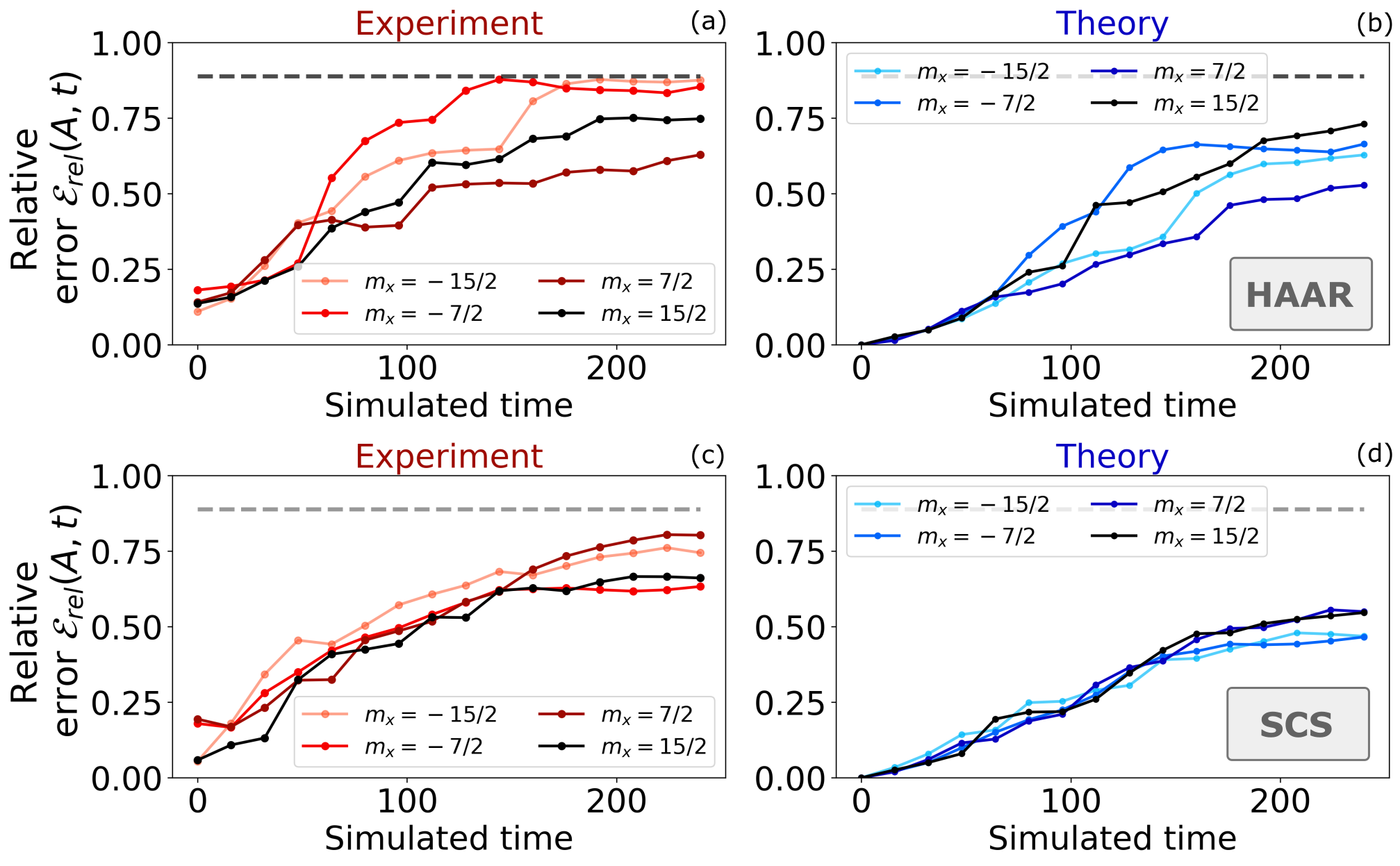}
	\caption{{\footnotesize Cumulative average relative error in expectation values, corresponding to: \textbf{(a)}, \textbf{(c)} experimental quantum simulation results and \textbf{(b)}, \textbf{(d)} numerical simulations from our proposed theoretical model. The shown data compliments Fig. 4 of the main text. Results shown are averages over 10 initial states, chosen randomly from \textbf{(a)-(b)} the uniform (Haar) distribution over Hilbert space and \textbf{(c)-(d)} the uniform distribution of spin coherent states. In all cases, the output observable is a projector $P_m$ onto a basis state of $S_x$. Dashed lines correspond to the theoretical prediction for the asymptotic value of the error $\mathcal{E}(A,\infty)$ (for Haar-random initial states). All data shown corresponds to simulation of the LMG evolution, for $B=0.4\Lambda$.}}
	
	\label{fig:fig3_extra}
\end{figure*}

\section{Additional numerical results} \label{app:num_extra}

In this section we present additional numerical results regarding the dynamics of the average relative error $\mathcal{E}_{rel}(A,t)$, defined in Eq. (9) of the main text. These results illustrate the generic dependence of this quantity on the observable purity, irrespective of the particular observable, model and system size we consider. In Fig. \ref{fig:fig_dim} we plot $\mathcal{E}_{rel}(A,t)$ for several choices of Hamiltonian, observable and type of initial states. All cases are shown for three values of particle numbers $N=2S=15,45,80$. We show cases corresponding to both the paramagnetic ($B>\Lambda$) and ferromagnetic ($B<\Lambda$) phases, and we use different sets of initial states like Haar-random states (HRS), spin coherent states (SCS) or Dicke states (DS). In all cases shown, the long time cumulative error becomes larger as the observable purity increases. We also observe good agreement between the numerical results (which are exact) and the analytical predictions for the long-time averages. Among all these cases, some differences are seen. For instance, the deviations from the mean are larger for SCS and DS with respect to the HRS. This behavior can be expected from the lack of structure of the HRS, which are prone to lead to self-averaging of the expectation value errors. For SCS and $s$ in the paramagnetic phase, we observe that the timescale in which saturation is achieved increases greatly, as can be seen in case \textbf{(d)}. We attribute this behavior to the combination of an integrable model with a very regular energy spectrum, and the choice of highly structured initial states.

\begin{figure*}[t]
	\centering
	\includegraphics[width=0.9\linewidth]{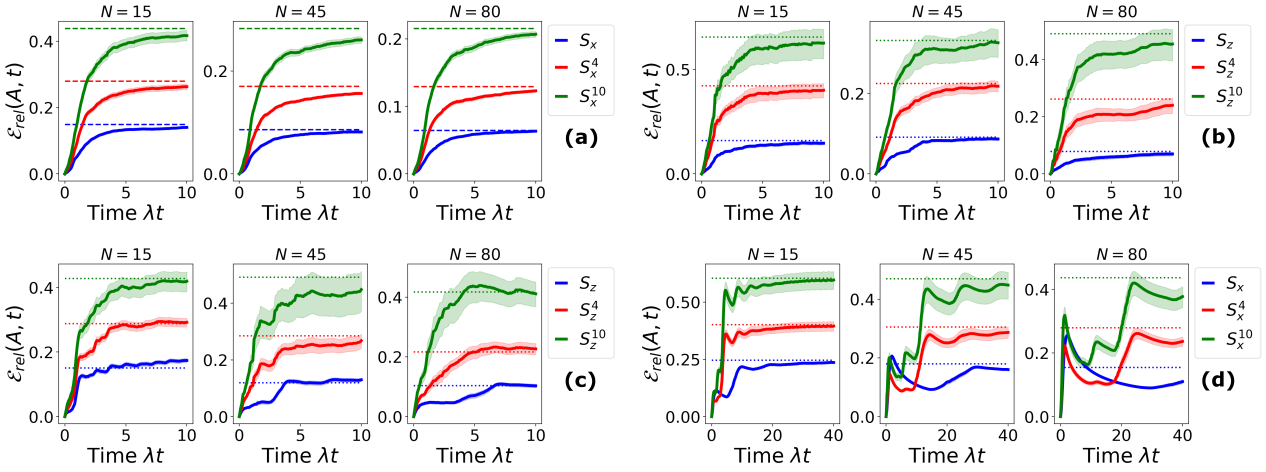}
	\caption{{\footnotesize Cumulative average relative error in expectation values, calculated from numerical simulations of the LMG dynamics with different values of system size $N=2S$. All cases show results for three observables chosen to have increasing purity $S_\alpha$, $S_\alpha^4$ and $S_\alpha^{10}$. In \textbf{(a)}-\textbf{(d)}, different cases are shown to illustrate the generic behavior of the relative error. \textbf{(a)} Initial states are Haar-random, $s=0.4$, $\alpha=x$. \textbf{(b)} Initial states are random Dicke states (i.e. eigenstates of $S_z$), $B=0.25\Lambda$, $\alpha=z$. \textbf{(c)} Initial states are random spin coherent states, $B=0.11\Lambda$, $\alpha=z$. \textbf{(d)} Initial states are random spin coherent states, $B=2.3\Lambda$, $\alpha=x$. For case \textbf{(a)}, dashed line correspond to the analytical prediction for the long-time value, c.f. Eq. (10) in the main text. For \textbf{(b)}-\textbf{(d)} such prediction is not available, so we plot as a dotted line the numerically computed average of the single-instance analytical results, Eq. (28) in the main text. In all cases, results correspond to averages over 50 initial states (shaded regions indicate standard errors). Each evolution, in turn, is obtained by averaging over 50 random perturbations, and the perturbation strength is set to $\lambda=0.025$ in all cases.   }}
	\label{fig:fig_dim}
\end{figure*}


\section{Application to other types of perturbations and systems} \label{app:other_systems}

Throughout this work we have analyzed the role of the observable purity in determining the average sensitivity of expectation values to deviations in the quantum state. As argued in the main text, we expect the core of our results to hold generically, irrespective of the details of the Hamiltonian and the perturbation. In this section we present a numerical analysis showing the applicability of some of our results to another paradigmatic many-body system, the transverse Ising model (TIM) \cite{suzuki2012}. The TIM describes the dynamics of $N$ spin-$\frac{1}{2}$ particles arranged in a one dimensional chain with nearest-neighbor interactions, and its Hamiltonian reads
\begin{equation}
H_{I} = -\frac{h}{2} \sum\limits_{i=1}^N \sigma_x^{(i)}- \frac{J}{4}\sum\limits_{i=1}^{N-1} \sigma_z^{(i)}\sigma_z^{(i+1)}.
\label{ec:tfim_hami}
\end{equation} 

In Fig. \ref{fig:tfim_plots} \textbf{(a)} we show plots of the cumulative relative error, averaged over $n_s=20$ Haar-random initial states and corresponding to the following choices of observables (for $N=6$)
\begin{eqnarray}
A_1 &=& \sigma_y^{(3)}\\
A_2 &=& \sigma_y^{(3)}\otimes\sigma_y^{(4)}\\
A_3 &=& \KetBra{\uparrow\ldots\uparrow}{\uparrow\ldots\uparrow}_{N-1}\otimes \mathbb{I}_N \\
A_4 &=& \KetBra{\uparrow\ldots\uparrow}{\uparrow\ldots\uparrow}
\label{ec:obs_ising}
\end{eqnarray}

Following the discussion in Appendix \ref{app:purity}, the purities of these operators satisfy $\eta(A_1)=\eta(A_2)<\eta(A_3)<\eta(A_4)$. As can be seen from the figure, the obtained results follow the same behavior as was observed for the LMG, where the average relative error is higher for higher purity observables.\\

The numerical simulations leading to Fig. \ref{fig:tfim_plots} \textbf{(a)} use as model for the random perturbation matrix $V$ a full Gaussian random matrix ensemble (the GOE). This same choice was used for all the LMG numerical calculations shown in this work. However, the TIM has a natural notion of spatial locality, which the LMG lacks. It is thus instructive to also analyze for this system the effects of imperfections arising from fluctuating local fields. For instance, consider the following perturbation Hamiltonian
\begin{equation}
V_{s} = \frac{1}{2}\sum\limits_{j=1}^N v_j \sigma_x^{(j)}
\label{ec:pert_vs}
\end{equation}

\noindent where we take the local fields to be normally distributed $v_j=\mathcal{N}(0,1)$. In Fig. \ref{fig:tfim_plots}  \textbf{(b)} we show results corresponding to the perturbation being of the form in Eq. (\ref{ec:pert_vs}). As can be observed from comparison with \textbf{(a)}, results are qualitatively similar, and the observable purity still plays a major role in determining the average relative error. We recall that the structure of the whole matrix $V$ does not enter the first order perturbation theory calculation that is described in the main text, and so it is not surprising that even a structured random matrix leads to qualitatively similar results (see also work in Ref. \cite{dabelow2020} and references therein).\\

\begin{figure}[t]
	\centering
	\includegraphics[width=1.\linewidth]{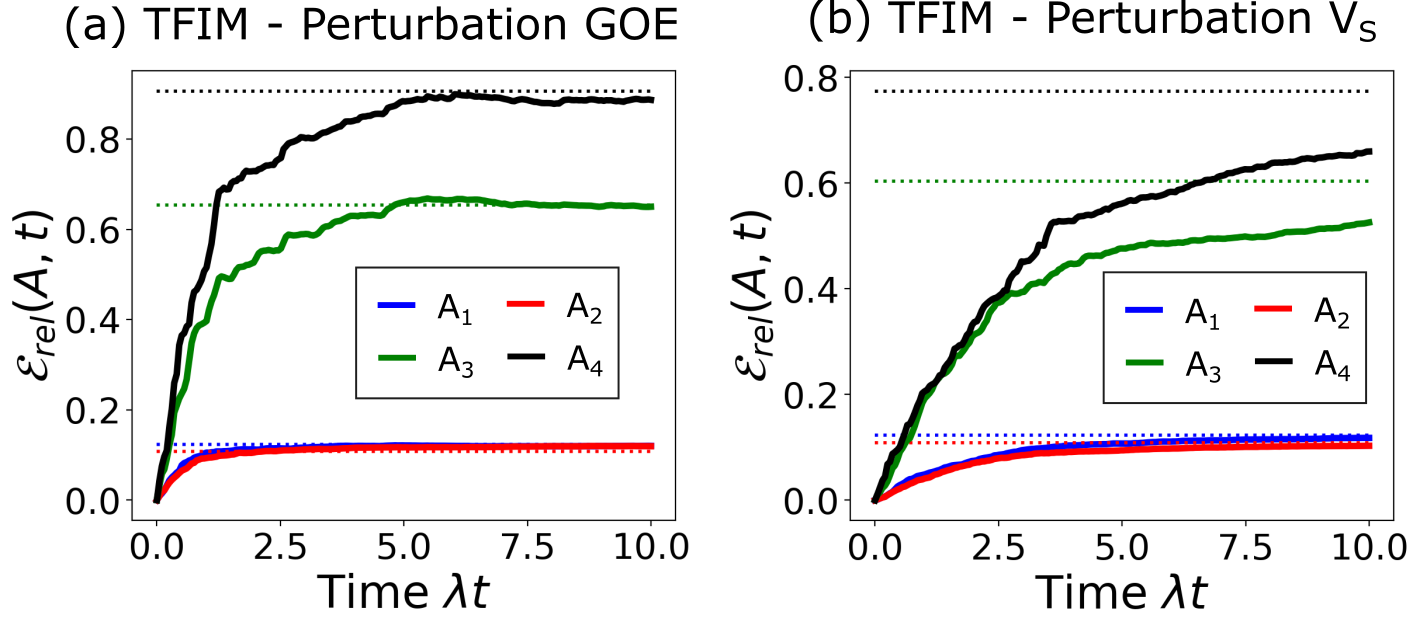}
	\caption{{\footnotesize Cumulative average relative error in expectation values, calculated from numerical simulations of the transverse Ising model with $N=6$ particles and $h/J=0.33$. All cases show results for four observables $A_i$ defined in Eq. (\ref{ec:obs_ising}), whose purities satisfy $\eta(A_1)=\eta(A_2)<\eta(A_3)<\eta(A_4)$. \textbf{(a)} Random perturbation Hamiltonian $V$ is drawn for the GOE ensemble, as done in all previous numerical results for the LMG model. \textbf{(b)} Random perturbation Hamiltonian $V$ is taken from the model in Eq. (\ref{ec:pert_vs}). Dashed lines show the numerically computed average of the single-instance analytical results, Eq. (28) in the main text. In can be readily observed that the overall behavior between both types of perturbation is very similar. The number of initial states here is $n_s=20$. Each evolution, in turn, is obtained by averaging over 50 random perturbations, and the perturbation strength is set to $\lambda=0.025$ in all cases.   }}
	\label{fig:tfim_plots}
\end{figure}

Finally, we point out that we expect noise channels with tensor product structure to determine differences in the sensitivity of same purity observables, particularly Pauli operators of different weights. Such differences are not prominent in the data shown in Fig. \ref{fig:tfim_plots}, as they are washed out by the initial Haar-random configurations. However, they are likely to appear for other types of initial states (such as Dicke or spin coherent states, which are separable) and/or for short times. We demonstrate this point in Fig. \ref{fig:tfim_plots2} where we show the cumulative error in expectation values for a initial separable state $\Ket{\psi_0}=\Ket{\uparrow\ldots \uparrow}_x$ as a function of time. The observables considered here are (for $N=8$)
\begin{eqnarray}
B_1 &=& \sigma_x^{(4)} \\
B_2 &=& \sigma_x^{(4)}\otimes\sigma_x^{(5)} \\
B_3 &=& \sigma_x^{(3)}\otimes\sigma_x^{(4)}\otimes\sigma_x^{(5)} \\
B_4 &=& \sigma_x^{(3)}\otimes\sigma_x^{(4)}\otimes\sigma_x^{(5)}\otimes\sigma_x^{(6)}
\label{ec:obs_ising2}
\end{eqnarray}

\noindent which all have the same purity as discussed previously. In case \textbf{(a)} we show the errors arising for the GOE perturbation model, for which no substantial differences are observed between these observables. On the other, with the local noise model shown in \textbf{(b)} we observe higher cumulative errors as the weight of the Pauli observable increases.

\begin{figure}[t]
	\centering
	\includegraphics[width=1.\linewidth]{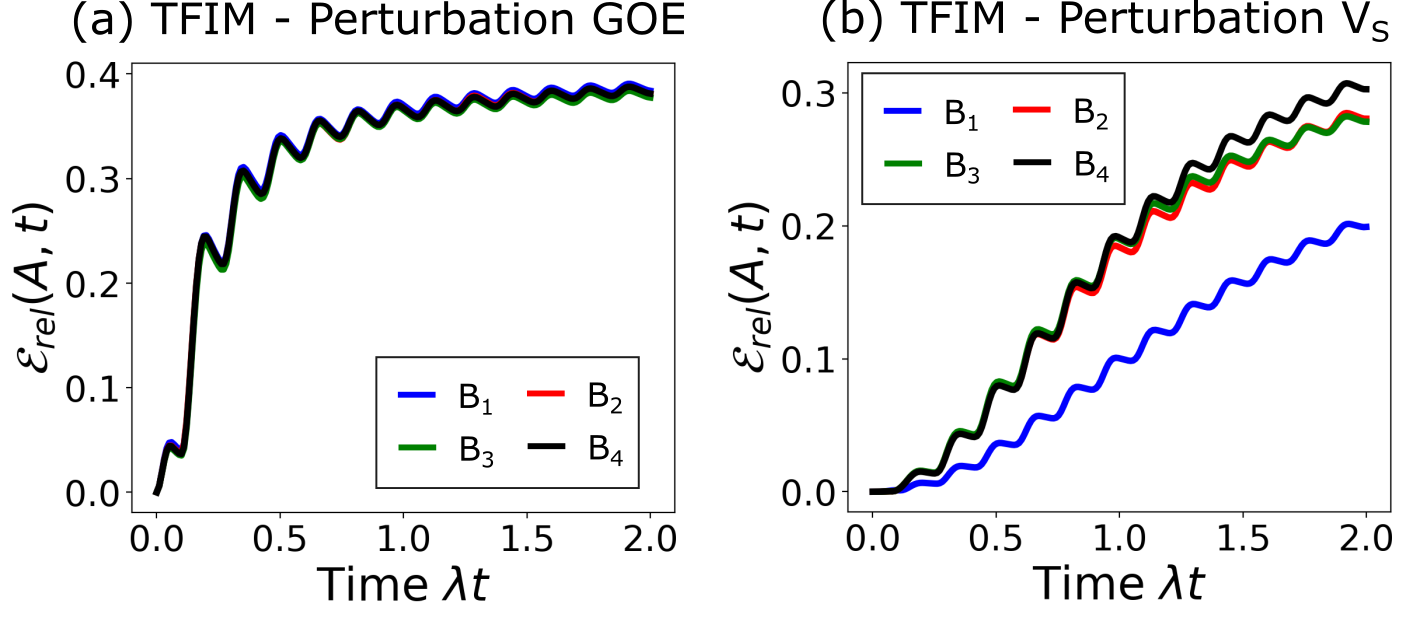}
	\caption{{\footnotesize Cumulative average relative error in expectation values, calculated from numerical simulations of the transverse Ising model with $N=8$ particles and $h=0.$. All cases show results for four observables $B_i$ defined in Eq. (\ref{ec:obs_ising2}). These are all Pauli observables (albeit of different weights), and so have the same observable purity. \textbf{(a)} Random perturbation Hamiltonian $V$ is drawn for the GOE ensemble. \textbf{(b)} Random perturbation Hamiltonian $V$ is taken from the model in Eq. (\ref{ec:pert_vs}). Results are shown for a particular initial state, $\Ket{\psi_0}=\Ket{\uparrow\ldots \uparrow}_x$. The evolution is obtained by averaging over 50 random perturbations, and the perturbation strength is set to $\lambda=0.025$ in all cases.   }}
	\label{fig:tfim_plots2}
\end{figure}

\bibliography{errors_aqs_ref} 

\end{document}